\documentclass[%
reprint,
 amsmath,amssymb,
 aps,
prx,
]{revtex4-2}
\usepackage[english]{babel}
\usepackage{graphicx}
\graphicspath{ {./maintext_images/} }
\usepackage[version=4]{mhchem} 
\usepackage{multirow}
\usepackage{amsmath,amsfonts,bm}
\usepackage[section]{placeins}
\usepackage{float}
\usepackage{epsfig}
\usepackage{fancybox}
\usepackage{subfigure}
\usepackage[dvipsnames]{xcolor}
\usepackage[titletoc]{appendix}
\usepackage{sidecap}
\usepackage[mathscr]{euscript}
\usepackage{physics}           % \bra, \ket, \braket, \mel, \abs, \norm
\usepackage{threeparttable}

\usepackage{soul}
\usepackage{siunitx}
\newcommand{\HH}{\ce{H2}}
\newcommand{\COO}{\ce{CO2}}
\newcommand{\NNO}{\ce{N2O}}

\newcommand{\cm}{cm$^{-1}$}
\newcommand{\pHH}{\textit{para}-\ce{H2}}
\newcommand{\oHH}{\textit{ortho}-\ce{H2}}
\newcommand{\kp}{\ensuremath{k_\mathrm{p}}}
\newcommand{\ko}{\ensuremath{k_\mathrm{o}}}

\newcommand{\Csixty}{\ce{C60}}

\newcommand{\avg}[1]{\left<#1\right>}

\newcommand{\kT}{\ensuremath{k_{\rm B}T}}

\renewcommand{\vb}{\ensuremath{\mathbf{v}}}

\newcommand{\kB}{k_{\mathrm{B}}}

\newcommand{\nB}{\bar{n}}

\begin{document}

\author{Nathan McLane}
\email{nmclane@umd.edu}
\affiliation{%
 Institute for Physical Science and Technology, \\ University of Maryland, College Park, MD 20742, USA
}%

\author{Richard C. Remsing}
\email{rick.remsing@rutgers.edu}
\affiliation{%
 Department of Chemistry and Chemical Biology, \\ Rutgers University, Piscataway, New Jersey 08854, USA}%

\author{Leah G. Dodson}
\email{ldodson@umd.edu}
\affiliation{%
 Department of Chemistry and Biochemistry, \\ University of Maryland, College Park, MD 20742, USA
}%

\date{\today}% It is always \today, today,

\title{Classical control of quantum-state relaxation in confined molecular hydrogen}

%\title{Molecular quantum-state dynamics governed by classical strain fields}

\preprint{arXiv:2610.XXXXX}

\begin{abstract}
Quantum-state conversion between the ortho and para nuclear-spin states of \HH{} is forbidden in the isolated molecule but these states can become connected through interactions with a surrounding environment. How that same environment governs the dynamics of the newly connected states remains poorly understood. Here, we investigate this problem using \HH{} confined in crystalline molecular solids, where the host environment selectively enables conversion between a pair of otherwise disconnected \oHH{} and \pHH{} states. Rotationally resolved infrared spectroscopy directly determines the state-specific conversion dynamics in crystalline \COO{} and \NNO{} at 10 K, revealing that increasing \HH{} concentration suppresses nuclear-spin conversion by nearly an order of magnitude, reaching complete suppression by 20\% \HH{} loading. Molecular dynamics and Monte Carlo simulations identify two complementary origins of this concentration dependence: changes in the thermally populated phonon modes available to mediate relaxation, and strain-driven \HH{} clustering that progressively removes molecules from the host crystal field responsible for enabling conversion. The resulting picture connects a classical collective response of the host to microscopic quantum-state dynamics: \HH{} loading generates strain that reorganizes the local molecular environment, thereby modifying both the phonon bath and the molecular coupling to that bath. These results establish a multiscale mechanism by which condensed-phase environments can govern the connectivity and dynamics of an effective two-level system.
  
\end{abstract}

\maketitle

\section{Introduction}

Open quantum systems are an active area of research with applications across a diverse range of fields, including quantum information and sensing, chemical dynamics, and machine learning \cite{mattes_designing_2025,ilias_criticality-enhanced_2022,bermudez_dissipation-assisted_2013,fallas_padilla_delocalized_2025,hartmann_neural-network_2019}. Broadly, an open quantum system consists of a localized microscopic subsystem coupled to an external environment \cite{rotter_review_2015,kapral_quantum_2015}. The dynamics of the microscopic subsystem are then governed by both its intrinsic behavior and its interaction with the environment. Therefore, to understand the intrinsic dynamics of the subsystem, the influence of the environment must be taken into account. In molecular and chemical physics, understanding the influence of solid-state environments on the quantum dynamics of confined molecular species is important for molecular quantum state preparation and preservation, as well as their potential application as qubits \cite{bader_confined_2009,bayliss_enhancing_2022,albert_robust_2020}. A particularly simple model for probing the influence of a confining environment is a system containing two otherwise disconnected quantum states whose connectivity is induced by interactions with the environment \cite{leggett_dynamics_1987,albert_robust_2020}. Molecular hydrogen in confinement provides an accessible target for this scenario. 

Molecular hydrogen (\HH{}), the simplest neutral molecule, has served as a foundational system for studies of molecular quantum mechanics over the last century \cite{wigner_uber_1933,farkas_orthohydrogen_1935,van_kranendonk_induced_1958,yildirim_rotational_2002,kohama_rotational_2009,ilisca_electron_2014}. The properties of isolated \HH{} have been characterized extensively through both experiment and theory, making it a particularly well-understood molecular system. Of particular relevance here are the two nuclear-spin states of \HH{}, \oHH{} (total nuclear spin $I=1$) and \pHH{} ($I=0$). Because the two protons are identical fermions, the total molecular wavefunction must be antisymmetric under proton exchange, correlating the rotational and nuclear-spin symmetries in the ground electronic and vibrational state \cite{herzberg_infrared_1946,herzberg_molecular_1950}. As a result, \oHH{} occupies odd-$j$ rotational states, whereas \pHH{} occupies even-$j$ states. 

In an isolated \HH{} molecule, transitions between the ortho and para manifolds are forbidden by symmetry; interactions with an external environment are therefore required to enable nuclear-spin conversion (NSC) by introducing new quantum-state connectivity \cite{herzberg_molecular_1950,silvera_solid_1980,curl_nuclear_1967,nielsen_paramagnetic_1967,ilisca_electron_2014}.  Confinement in an asymmetric environment can break the molecular symmetry and transform the isolated-molecule problem into a broken-symmetry many-body problem in which the quantum-state properties and dynamics of \HH{} are modified by interactions with the surrounding medium. Previous studies have shown that confinement can lift the degeneracy of the magnetic quantum number $m$ sublevels within the $j=1$ (ortho) manifold \cite{schweizer_nmr_1979,fitzgerald_quantum_2008,di_cataldo_giant_2024}. In our recent work, we further found that conversion to the $j=0$ (para) state occurs selectively from the $j=1, m=0$ state in certain crystalline molecular solids \cite{mclane_environment-imposed_2026}. The resulting conversion dynamics are therefore restricted to an effective two-level system comprising the $j=1, m=0$ ortho state and the $j=0$ para state. Previously, we used the $\ket{\lambda m}$ notation with $\lambda = 1$ and 0 for ortho and para states, respectively, where $\lambda$ reflects the total angular momentum in the coupled translational--rotational system; here, we simplify the notation to $\ket{jm}$ as $\lambda=j$ under these conditions. This environment-defined state connectivity between $\ket{00}$ (para) and just the $\ket{10}$ ortho substate suggests a route toward isolating and controlling molecular two-level systems relevant to quantum-state manipulation. 

\begin{figure}[h]
    \centering
     \includegraphics[width=3.25in]{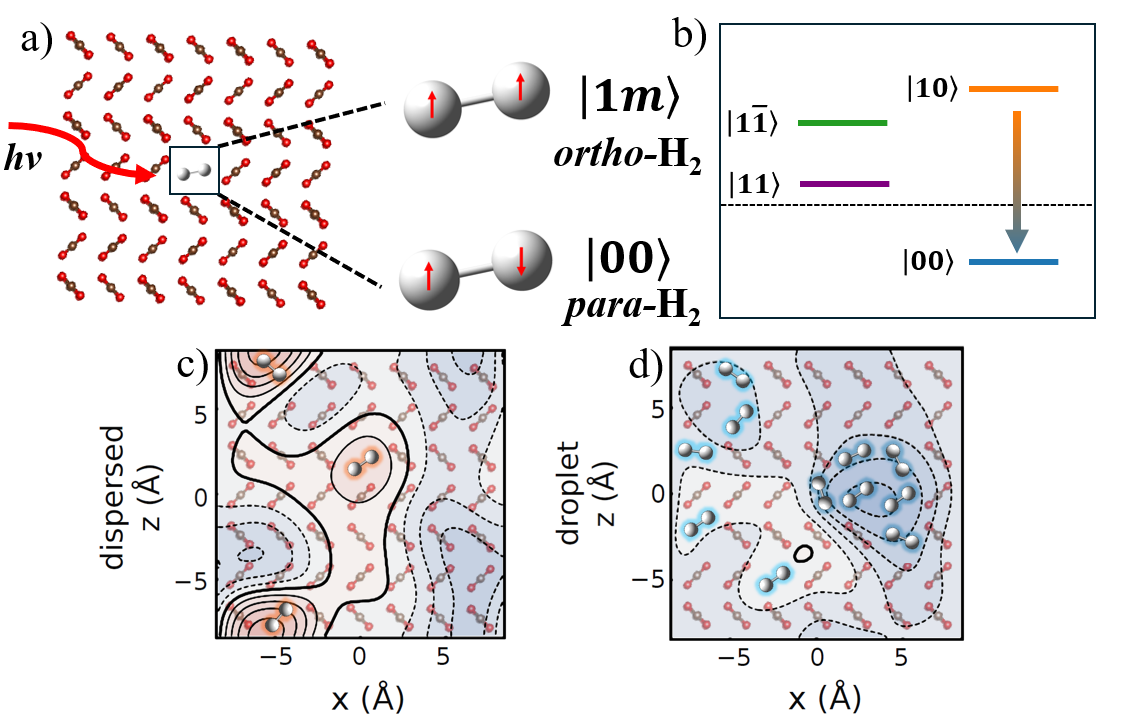}% 
    \caption{Microscopic quantum dynamics of \HH{} confined in crystalline \COO{}. a) Infrared spectroscopy resolves individual states (ortho/para) and substates (ortho $|m|=0,1$) of molecular hydrogen confined in crystalline \COO{}. b) Splitting of the correlated rotational states in the \oHH{}, $\ket{1m}$ manifold and environment-imposed selection rules for the transition from $\ket{10}$ (ortho) to $\ket{00}$ (para). c) Dispersed \HH{} in a compressive volumetric strain field relaxing into d), a clustered arrangement with reduced strain.}  
    \label{fig:overview}
\end{figure}

Here, we investigate how the properties of the confining environment control the dynamics of nuclear-spin conversion within this environment-defined state space. Using rotationally resolved infrared spectroscopy, we directly monitor the population of each state ($\ket{10}$ ortho and $\ket{00}$ para) as a function of time in the effective two-level system created by the surrounding crystalline molecular solid (Fig. \ref{fig:overview}(a) and Fig. \ref{fig:overview}(b)). We find that increasing the \HH{} concentration substantially slows the nuclear-spin conversion, revealing the guest concentration itself as a control parameter for the quantum-state dynamics. The trend persists across both \COO{} and \NNO{} hosts, demonstrating that the concentration dependence is not unique to a specific host. Molecular dynamics and Monte Carlo simulations connect this concentration dependence to changes in both lattice dynamics and the spatial distribution of \HH{} within the host. With low \HH{} concentrations, the relaxation rates are largely dictated by phonon-induced nuclear-spin relaxation that diminishes in importance as the phonon density of states of the host shifts to higher frequencies with increased \HH{} loading. At higher concentrations, strain within the host promotes \HH{} clustering, which emerges as the dominant contribution to the suppression of nuclear-spin conversion (Fig. \ref{fig:overview}(c) and Fig. \ref{fig:overview}(d)). The resulting picture connects a classical collective response of the host to the quantum dynamics of the confined molecule: \HH{} loading generates strain that reorganizes the microscopic environment responsible for enabling nuclear-spin conversion in the first place.

We proceed in Sec. \ref{sec:expt} with a description of the rotationally resolved infrared spectroscopy used to monitor nuclear-spin conversion of confined \HH{}. In Sec. \ref{sec:theor}, we introduce the molecular dynamics and Monte Carlo simulations used to characterize lattice dynamics and the spatial organization of \HH{} within the host. In Sec. \ref{sec:expt-res}, we quantify the concentration-dependent nuclear-spin conversion kinetics and develop a phonon-mediated description of the observed relaxation. In Sec. \ref{sec:theor-res}, we connect these dynamics to concentration-induced strain and \HH{} clustering, establishing the relationship between the collective response of the host and the quantum dynamics of the confined molecule. In Sec. \ref{sec:discussion}, we discuss the implications of this work in establishing environment-defined effects on the quantum dynamics of confined subsystems.

\section{Rotationally Resolved Spectroscopy of Nuclear-Spin Conversion} \label{sec:expt}

Because the nuclear-spin and rotational symmetries of \HH{} are correlated, rotationally resolved spectroscopy provides a direct probe of the nuclear-spin populations: odd-$j$ rotational states report the \oHH{} population, whereas even-$j$ states report \pHH{}. However, in the isolated \HH{} molecule the absence of a permanent dipole moment imposes restrictive selection rules that make these spectroscopic transitions formally forbidden \cite{herzberg_molecular_1950,herzberg_infrared_1946}. Confinement perturbs the molecular symmetry and induces interactions with the surrounding medium that relax these selection rules, providing sufficient intensity to resolve the rotational-state populations directly by rotational--vibrational infrared spectroscopy \cite{fitzgerald_quantum_2008,van_kranendonk_induced_1958,chisholm_induced_1954}. We exploit this environment-induced spectroscopic accessibility to monitor nuclear-spin conversion as a function of time.

The infrared (IR) spectroscopy instrument has been described in previous work \cite{hockey_weakly_2022,hockey_matrix-formation_2024,mclane_environment-imposed_2026}. Briefly, a mixture of gases is deposited on a ZnSe substrate that is mounted at the base of a CTI Model 22 cryostat that can be cooled to as low as 8 K. A Lakeshore model 331 temperature controller monitors and controls the temperature of the substrate via a silicon diode and resistive heaters. The cryostat extends through a differentially pumped rotary platform that enables the attached substrate to rotate between a ``deposition axis,''---where a flow of gaseous species collide with the cold substrate at normal incidence---and the ``spectroscopy axis,'' where frozen species are probed using IR light in transmission mode. 

In the present work, mixed ices form directly on the cold substrate by flowing gas mixtures into the chamber from prepared cylinders containing all species. Molecular hydrogen (\HH{}) concentrations (mole fractions) of 1\%, 5\%, 10\%, 15\% and 20\% were prepared by diluting pure \HH{} (Airgas, UHP) with excess CO$_2$ (Airgas, Industrial Grade) or N$_2$O (Airgas, USP) in a stainless-steel cylinder to total pressures of 1500 Torr, manometrically. The cylinder containing the gas mixture is attached to the vacuum chamber through a gas manifold system allowing deposition on the cold ZnSe substrate held at 10 K. Deposition rates are controlled through a VAT dosing valve, at a rate of 7500 mtorr L s$^{-1}$ for a total deposition time of 30 min. Spectra are taken after deposition and every 5 minutes for a total of at least 75 minutes.

IR spectra are obtained using a Nicolet iS50 Fourier-transform infrared spectrometer. The IR light makes a single pass through the combined ice and ZnSe optic and is detected by an external MCT-A detector. The spectra are recorded from 5000 to 600 cm$^{-1}$ with a resolution of 0.5 cm$^{-1}$ and a data spacing of 0.241 cm$^{-1}$. 

The environment-induced spectral separation of the rotational states allows the population dynamics of the converting ortho and para states to be followed independently in the IR spectrum. Each time-dependent spectrum is fit in the fundamental rotational--vibrational transition region of \HH{} (Q$_1$, near 4160 cm$^{-1}$) to a linear combination of Gaussian functions to obtain the individual contributions of each \HH{} state at that reaction time. The center frequency and width of the function are obtained by using a global fit of the data set, and are subsequently fixed, allowing only the amplitude of the function to change as a function of time. The same spectra are also analyzed by obtaining a difference spectrum $I_\text{diff}(t) = I(t) - I(0)$, where $I(t)$ is the spectrum at time $t$ and $I(0)$ is the first spectrum taken immediately after deposition ends (assumed to be $t=0$). The difference spectrum is fit to a linear combination of Gaussian functions in the same fashion as the regular fitting mechanism, again obtaining areas that correspond to each \HH{} state at that time point, but reflecting now the change in population over the course of the reaction. Once properly fitted, the associated Gaussians are integrated as a function of time. The integrated areas from the difference spectra ($I_{\mathrm{diff},\alpha}(t)$) for each species $\alpha$ are plotted as a function of time and fit to a mono-exponential function to extract a state-specific relaxation rate:
\begin{equation}
     I_{\mathrm{diff},\alpha}(t) = (I_{\alpha}(0) - I_{\alpha}({\infty}))\exp({-k_{\alpha}t}) + I_{\alpha}({\infty})
     \label{eq:integ_int}
\end{equation}
where $I_{\alpha}(0)$ and $I_{\alpha}(\infty)$ are fitted parameters that correspond to the $t=0$ and $t=\infty$ intensities. Here $k_{\alpha}$ is the observed first-order relaxation rate in inverse seconds associated with the time evolution of state $\alpha$ (either $\ket{00}$ or $\ket{10})$. Error bars are calculated by propagating the error obtained by doing a global fit on the dataset to obtain a fixed line center and width, the fit to the experimental data, through the final NSC kinetics fit ($I$ vs. $t$). The largest component to the determined error is attributed to the determination of the line center and width resulting in similar magnitude error bars for each integrated data point. Agreement between independently extracted ortho-loss and para-growth rates provides an internal measure of whether the observed spectral evolution reflects population transfer between the two nuclear-spin manifolds.

\section{Molecular Modeling of the Confining Environment} \label{sec:theor}

The experimentally observed concentration dependence requires a mechanism by which changing the number of confined \HH{} molecules modifies either the environment-induced coupling between nuclear-spin states or the dynamics of the surrounding lattice. We therefore consider two complementary contributions to the conversion rate: the phonon spectrum of the host, which determines the available lattice vibrations, and the local crystal field experienced by \HH{}, which determines the strength of its coupling to those fluctuations.

In cryogenic crystals the induction of NSC of a species is mediated by spin--spin, spin--rotation, quadrupolar, and phonon-mediated processes \cite{miyamoto_nuclear_2008,mclane_environment-imposed_2026,fedders_nuclear-spin-lattice_1979,ebner_theoretical_1975,fedders_low-temperature_1986,fedders_spin-resonance_1975}.
Given our experimental temperatures (10~K) and \HH{} generally lacking the spin--rotation and spin--spin couplings, phonon processes can be expected to initiate the NSC observed in our system \cite{miyamoto_nuclear_2008}. 
Spin relaxation, in general, can be described by three processes: the one-phonon/direct process, and two two-phonon processes---the Orbach process and the Raman process \cite{miyamoto_nuclear_2008,abragam_principles_1961,scott_spin-lattice_1962,orbach_spin-lattice_1961}. 
When the energy difference between the rotational states $j=0$ and $j=1$ is much larger than the thermal energy, $\hbar\omega_{01}\gg \kT$, the contribution from the direct process is much smaller than that from two-phonon processes and can be treated as a constant \cite{miyamoto_nuclear_2008,abragam_principles_1961}. 
The Orbach process is relevant when the energy difference between the initial and some intermediate state of \HH{} is smaller than $k_{\rm{B}}T_{\rm{D}}$ where $T_{\rm{D}}\approx 120$~K is the Debye temperature for \HH{} \cite{scott_spin-lattice_1962}. 
At the low temperature considered here, only the $j=0$ and $j=1$ states of \HH{} are populated, with a large gap between the $j=1$ and $j=2$ states $(339.4$ K), such that the contribution from the Orbach process is negligible. 
As a result, we anticipate that the Raman process dominates NSC.

The rotational transition is driven by a two-phonon Raman process, in which a phonon of wavevector $\bm{k}$
is absorbed and one of wavevector $\bm{k}'$ is emitted, with energy
difference $\hbar(\omega_{\bm{k}} - \omega_{\bm{k}'}) = \hbar\omega_{nn'}$.
The transition rate from state $n$ to state $n'$ is
\begin{align}
  \Gamma^{(2)}_{n \to n'} &= \frac{2\pi}{\hbar^2} \sum_{\bm{k},\bm{k}'}
  \bigl|V^{(2)\,nn'}_{\bm{k}\bm{k}'}\bigr|^2 \; \nonumber \\
  &\times \nB(\omega_k)\bigl[\nB(\omega_{k'})+1\bigr] \;
  \delta(\omega_k - \omega_{k'} - \omega_{nn'}),
  \label{eq:Gamma_Raman}
\end{align}
where $V^{(2)\,nn'}_{\bm{k}\bm{k}'}$ is the effective matrix element at second order in the phonon--rotor coupling and $\nB(\omega) = (e^{\hbar\omega/\kB T}-1)^{-1}$ is the Bose--Einstein distribution  \cite{abragam_principles_1961}.
We can convert the sums over wave vectors to integrals by introducing the phonon density of states, $g(\omega)$.
Upon integrating over the delta function, and assuming that $\omega_{nn'}=\omega_n-\omega_{n'}$ is significantly smaller than $\omega_n$ and $\omega_{n'}$ except for in a negligible fraction of the phonon spectrum, such that $\omega_n=\omega_{n'}$ or $\omega_{nn'}=0$, we can estimate the transition rate as
\begin{equation}
  \Gamma^{(2)}_{n\to n'} = A_{nn'} \int g(\omega)^2\;
  \nB(\omega)\bigl[\nB(\omega)+1\bigr]\; \mathrm{d}\omega,
  \label{eq:Raman_zero}
\end{equation}
where 
\begin{equation}
  A_{nn'} \propto \left|\mel{\psi_n}{\frac{\partial V_{\mathrm{CF}}}{\partial u}}{\psi_{n'}}\right|^2
  \label{eq:coupling_constant}
\end{equation}
is a coupling constant that absorbs the squared phonon--rotor coupling matrix element,
$V_{\rm{CF}}$ is the interaction potential or crystal field on the H$_2$ molecule from the surrounding molecules in the crystal \cite{miyamoto_nuclear_2008}, $\left | \psi_n\right>$ is a rotational eigenstate, and $u$ is the displacement of the host molecules from their equilibrium lattice positions relative to the \HH{} center of mass, expanded in phonon normal modes, and the derivative is evaluated at the equilibrium configuration.
For a Debye solid at low temperatures, the rate predicted by Eq.~\ref{eq:Raman_zero} scales as $T^7$ \cite{miyamoto_nuclear_2008,scott_spin-lattice_1962}.
Eq. \ref{eq:Raman_zero} identifies two distinct routes by which \HH{} concentration can modify the NSC dynamics: concentration may alter the phonon density of states, $g(\omega)$, or it may alter the local crystal field and therefore the coupling prefactor, $A_{nn'}$. These two contributions provide a framework for connecting the macroscopic composition of the solid to the microscopic quantum-state dynamics of \HH{}. 
Because $V_{\rm CF}$ is determined largely by local quadrupole--quadrupole interactions, $A_{nn'}$ should not change much with \HH{} concentration at concentrations where the \HH{} molecules are randomly distributed or dispersed throughout the crystal.
In contrast, the density of states, $g(\omega)$, can be expected to change because the substitution of \COO{}
%or \NNO{} 
with \HH{} should disrupt long-wavelength phonons by serving as a phonon scattering center. 
Within this quantum theory for spin relaxation, we propose that \HH{}-induced changes in the phonon density of states leads to a decrease in spin-conversion rates with increasing \HH{} concentration.
The phonon density of states can be computed  as a Fourier transform of the velocity autocorrelation function obtained from molecular dynamics (MD) simulations,
\begin{equation}
g(\omega) = \int_0^\infty dt \avg{\vb(t)\cdot \vb(0)} e^{-i\omega t},
\end{equation}
where $\avg{\vb(t)\cdot \vb(0)}$ is the velocity autocorrelation function, which is implicitly summed over all atoms and averaged over multiple time origins to improve statistics in an equilibrium simulation.
We computed densities of states, $g(\omega)$, for CO$_2$ systems from MD simulations to test the above hypothesis.
Initial configurations of H$_2$ in crystalline CO$_2$ were generated by substituting a randomly chosen CO$_2$ molecule with an H$_2$ molecule until the desired H$_2$ concentration was reached. 
MD simulations were performed using the Atomic Simulation Environment (ASE)~\cite{ASE} at 10~K in the microcanonical ensemble (constant $N$, $V$, and $E$) after extensive equilibration at constant temperature. 
Vibrational densities of states (VDOS) were computed by averaging over at least 20 independent NVE trajectories 100~ps in length.
The Orb universal neural network potential was used to describe the interatomic interactions in all the simulations~\cite{Orb,MarklandOrb}, which predicts the interatomic forces of H$_2$ in CO$_2$ at the level of a hybrid meta GGA density functional approximation (Fig. S6). 
The preceding analysis assumes that \HH{} remains randomly dispersed as its concentration increases. Because changes in the local \HH{} coordination would directly modify the crystal field contribution $A_{nn'}$, we independently test this assumption using replica exchange Monte Carlo (MC) simulations~\cite{swendsen_replica_1986,hukushima_exchange_1996,sugita_replica_1999,grigera_fast_2001,ninarello_models_2017}, which attempt to exchange the position of an H$_2$ molecule and a nearest neighbor molecule. 
MC moves were accepted or rejected using the Metropolis criterion, and the MC simulations were performed until the configuration of the system was equilibrated. 
These MC simulations enable us to determine if the H$_2$ molecules cluster or if they remain dispersed (or unclustered) in the crystals, especially at high H$_2$ concentrations. 

\section{Concentration-Dependent Nuclear-Spin Conversion} \label{sec:expt-res}

\begin{figure}[t]
    \centering
     \includegraphics[width=3.25in]{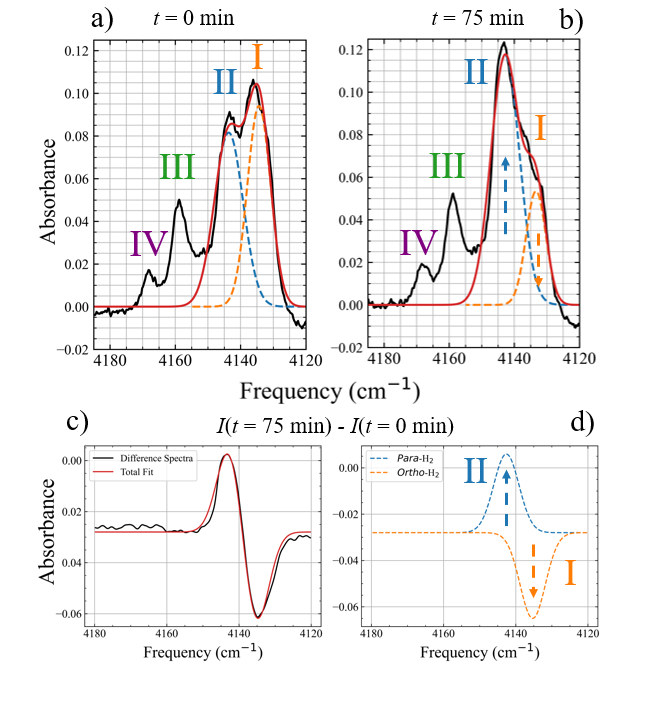}% 
    \caption{The Q$_1$ spectral region of 1\% \HH{} confined in \COO{} at 10 K. Linear absorbance spectra are plotted in the top row at two time points (a), $t = 0$; (b), $t = 75$ min. Data are shown in black and the two Gaussian fits for peaks I (orange) and II (blue), with the linear combination (red) are included. (c) The difference spectrum $I(t = 75 \text{ min}) - I(t = 0 \text{ min})$ with data in black and the linear combination of the two-Gaussian fit in red. (d) The two Gaussian fits for peaks I and II obtained from the difference spectrum, with growth of peak II (\pHH{}, blue) and loss of peak I (\oHH{}), orange.}  
    \label{fig:CO2}
\end{figure}

Rotationally resolved time-dependent spectra directly reveal state-selective population transfer between the \textit{ortho}- and \pHH{} manifolds. For 1\% \HH{} in \COO{} at 10~K, only one of the three spectral features assigned to \oHH{} evolves with time, and its loss is accompanied by growth of the \pHH{} feature. Comparing the background-subtracted linear absorbance spectra for 1\% \HH{} in \COO{} (Fig.~\ref{fig:CO2}(a) and Fig.~\ref{fig:CO2}(b)) at two representative time points, $t=0$ and $t=75$ min, shows that peak I, centered at 4134.4~\cm{} and assigned to Q$_1$ $\ket{1m}\rightarrow\ket{1m}$ transitions of \oHH{} where $\Delta m=0$ \cite{mclane_environment-imposed_2026}, decreased with a corresponding increase of peak II centered at 4143.9 \cm{}, the sole \pHH{} transition. Peaks III and IV at 4158.1 and 4168.2 \cm{} do not change with time. The time evolution for peaks I and II is most obvious in the difference spectra (Fig.~\ref{fig:CO2}(c)), and the time independence of peaks III and IV is especially clear as a flat line in the difference spectrum. Individual Gaussian fits of peaks I and II and linear combination spectra of rotationally resolved vibrational spectra are also shown for comparison (Fig.~\ref{fig:CO2}(d)). 

\begin{figure}[h]
    \centering
    \includegraphics[width=3.25in]{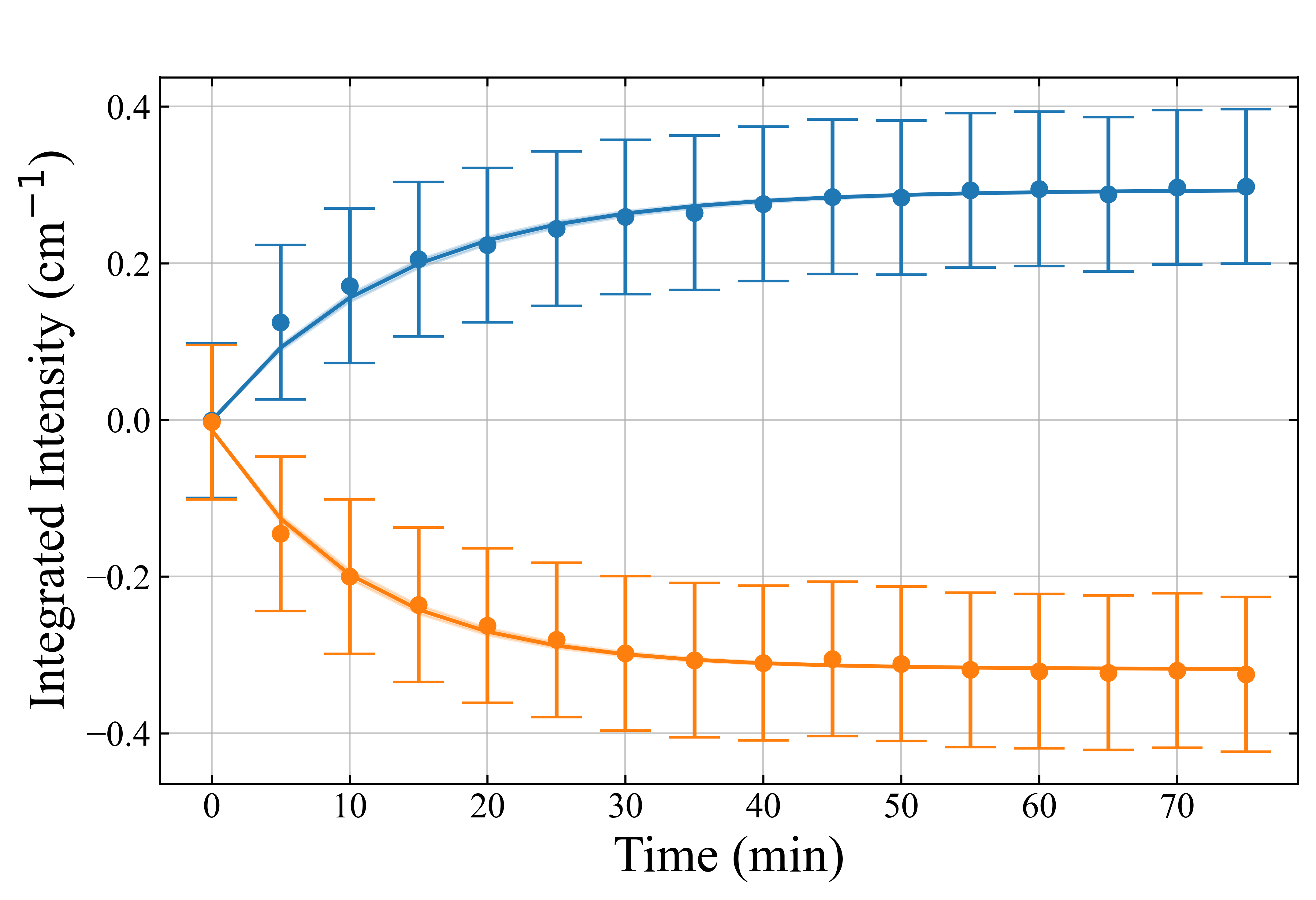}% 
    \caption{Time-dependent integrated Gaussian areas of the difference spectra intensities of peak II (\pHH{}, blue), peak I (\oHH{}, orange) for 1\% \HH{} in \COO{}. Solid lines represent the fit of Eq. \ref{eq:integ_int} to the integrated intensities as a function of time.}
    \label{fig:CO2_kinetics}
\end{figure}

The relaxation rates, corresponding to loss of \oHH{} and growth of \pHH{}, are obtained from the time-dependent evolution of peaks I and II. The time dependence of the integrated Gaussian areas for these peaks for all time points from 0 to 75 min is shown in Fig.~\ref{fig:CO2_kinetics} for the 1\% case. Taken from the difference spectra, the integrated intensities start at 0 \cm{} for each peak and then decay exponentially for \oHH{} ($\ket{10}\rightarrow\ket{00}$) with exponential growth mirrored for \pHH{}. The data is fit using Eq.~\ref{eq:integ_int} to obtain the relaxation rates using data from both the linear absorption spectra and the difference spectra and were consistent between the two methods. For example, the 1\% \HH{} in \COO{} data yielded relaxation rates for the linear absorption spectra of \SI{1.5e-3}{\per\s} and \SI{1.4e-3}{\per\s} for para growth and ortho loss, respectively, while fits using the difference spectra were \SI{1.3(0.6)e-3}{\per\s} and \SI{1.4(0.8)e-3}{\per\s} for the same conditions. The rest of this paper will discuss only the difference spectra results since the linear absorbance spectra gave consistent results. Fitted spectra and kinetics profiles for \HH{} concentrations from 5\% to 15\% are shown in the Supplemental Material (SM), Figs. S1--S3 \cite{SM}. 

\begin{table}[h]
    \caption{Summary of observed relaxation rates (\SI{e-3}{\per\s}) of \HH{} in \COO{} ices, measured for \pHH{}-gain ($\kp$) and \oHH{}-loss ($\ko$).}
    \label{tbl:rates}
    \centering
    \renewcommand{\arraystretch}{1.15}
    \begin{tabular}{c c|c|c|c}
        \hline
        \multirow{2}{*}{Relaxation rate} & 
        \multicolumn{4}{c}{\HH{} in \COO{}} \\
        %\cline{2-5}
         & 1\% \HH{} & 5\% \HH{}  & 10\% \HH{} & 15\% \HH{} \\
        \hline
        \kp  & 1.3 $\pm$ 0.6 & 0.4 $\pm$ 0.1 & 0.7 $\pm$ 0.2  & 0.2 $\pm$ 0.1  \\
        \ko  & 1.4 $\pm$ 0.8 & 0.5 $\pm$ 0.2 & 0.6 $\pm$ 0.2 & 0.3 $\pm$ 0.2 \\
        \hline
    \end{tabular}

\end{table}

The matched loss and growth rates provide direct evidence that the observed spectral evolution represents population transfer from the $\ket{10}$ ortho state into the $\ket{00}$ para state. This correspondence persists across all investigated \HH{} concentrations (Table~\ref{tbl:rates}), demonstrating that concentration modifies the rate of conversion without changing the state-selected connectivity established previously. We showed previously that peak I has contributions from \oHH{} molecules in different translation--rotation angular momentum states ($m=0,\pm1$) but that only the state with $m=0$ is allowed to convert to \pHH{}. Therefore, the peak I intensity decreases with time as the $\ket{10}$ state is depopulated, but it plateaus at a non-zero intensity as non-connected $|m|=1$ ortho states remain. Peaks III and IV were not observed to change with time in experiments that lasted out to 6 hours in any of the conditions used here; therefore, no relaxation rates are reported for these peaks that contain only contributions from the non-relaxing $|m|=1$ \oHH{} states. 

\begin{figure}[h]
    \centering
    \includegraphics[width=3.25in]{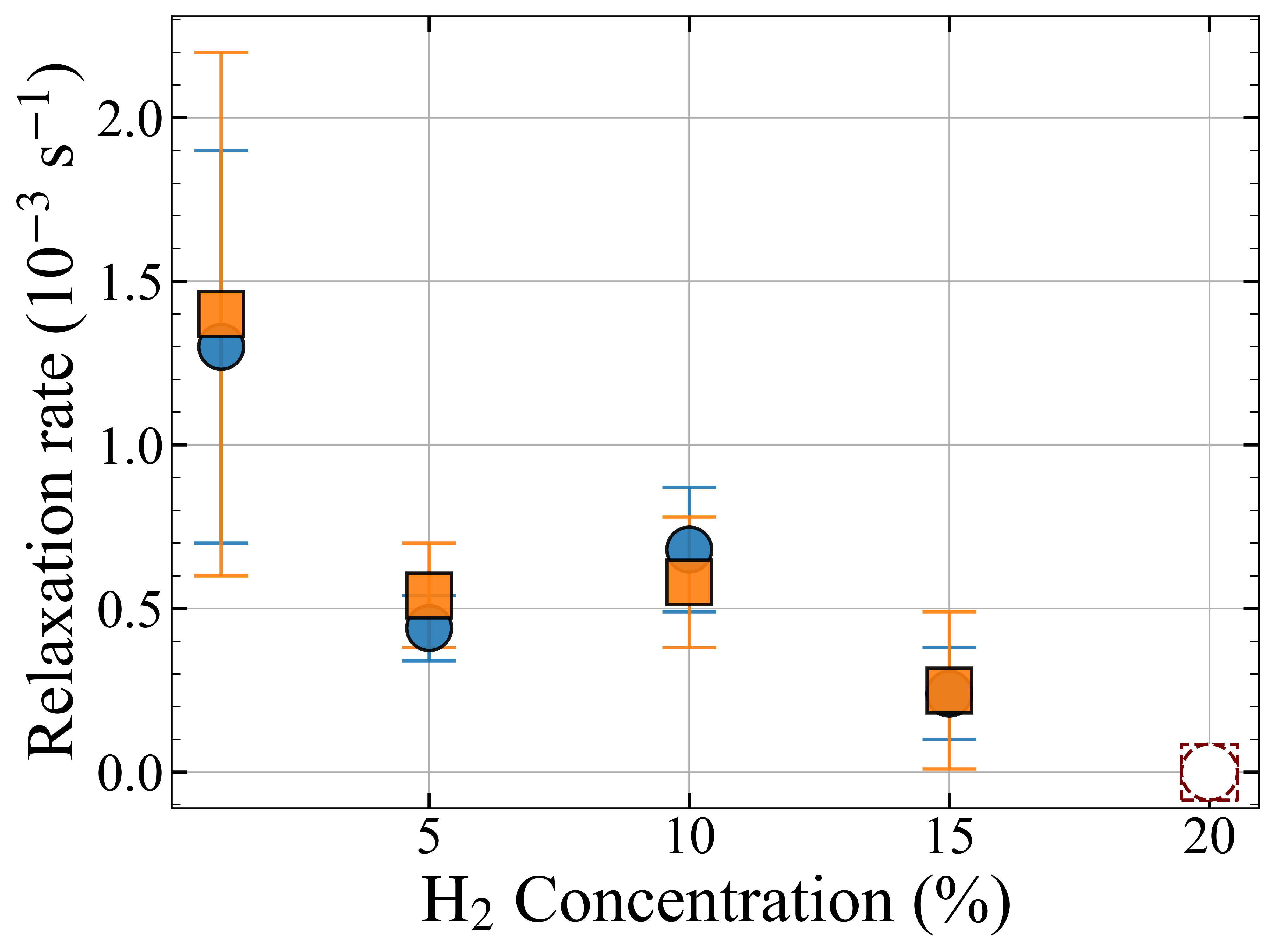}% 
    \caption{ Relaxation rate (\SI{e-3}{\per\s}) of \pHH{} (blue) and \oHH{} ($\ket{10}$) (orange) as a function of the total concentration of \HH{} (\%) in the deposited \COO{} ice mixture at 10 K. In experiments the relaxation rates at 20\% \HH{} were immeasurably slow, as indicated by the open symbols at 0 s$^{-1}$.}
    \label{fig:all_kinetics_plot}
\end{figure}

The central experimental result is a pronounced suppression of NSC with increasing \HH{} concentration. As shown in Fig.~\ref{fig:all_kinetics_plot}, the matched ortho-loss and para-growth rates decrease by nearly an order of magnitude between 1\% and 15\% \HH{} in \COO{}. At 20\% \HH{}, no measurable conversion is observed over six hours, indicated by the open symbols in Fig.~\ref{fig:all_kinetics_plot}. Thus, changing the concentration of the quantum subsystem itself strongly modifies the dynamics enabled by the surrounding host.

To test whether the concentration dependence is specific to the \COO{} host, we performed complementary measurements in crystalline \NNO{}. Spectra for \HH{} in \NNO{} are presented in the SM, along with the time-evolving intensities for peaks assigned to \oHH{} and \pHH{} (Figs.~S4--S5). Relaxation rates were obtained from mono-exponential fits; the results are listed in Table~\ref{tbl:rates-N2O} for the two sets of conditions investigated (1\% and 10\% \HH{} in \NNO{}). Again the relaxation rate for the ortho and para states is in close agreement with each other in tested conditions, in line with the expectation that \oHH{} in the $\ket{10}$ state converts to \pHH{} via the environment-induced connectivity in this crystalline host. Compared with the \COO{} case, the relaxation at 1\% \HH{} in \NNO{} is about three times as fast as the same conditions in \COO{}, but when the concentration of \HH{} reaches 10\%, the relaxation rates are similar between \COO{} and \NNO{} ices.

\begin{table}[h]
    \centering
    \caption{Summary of observed relaxation rates (\SI{e-3}{\per\s}) of \HH{} in \NNO{} ices, measured for \pHH{}-gain ($\kp$) and \oHH{}-loss ($\ko$).}
    \label{tbl:rates-N2O}
    \renewcommand{\arraystretch}{1.15}
    \begin{tabular}{c c|c}
        \hline
        \multirow{2}{*}{Relaxation rate} & 
        \multicolumn{2}{c}{\HH{} in \NNO{}} \\
         & 1\% \HH{} & 10\% \HH{} \\
        \hline
        \kp  & 3 $\pm$ 1 & 0.6 $\pm$ 0.5 \\
        \ko  & 3 $\pm$ 2 & 0.6 $\pm$ 0.1 \\
        \hline
    \end{tabular}

\end{table}

\section{Phonon-Induced Nuclear-Spin Dynamics and Strain-Driven Clustering} \label{sec:theor-res}

\begin{figure}[h]
\begin{center}
\includegraphics[width=0.4\textwidth]{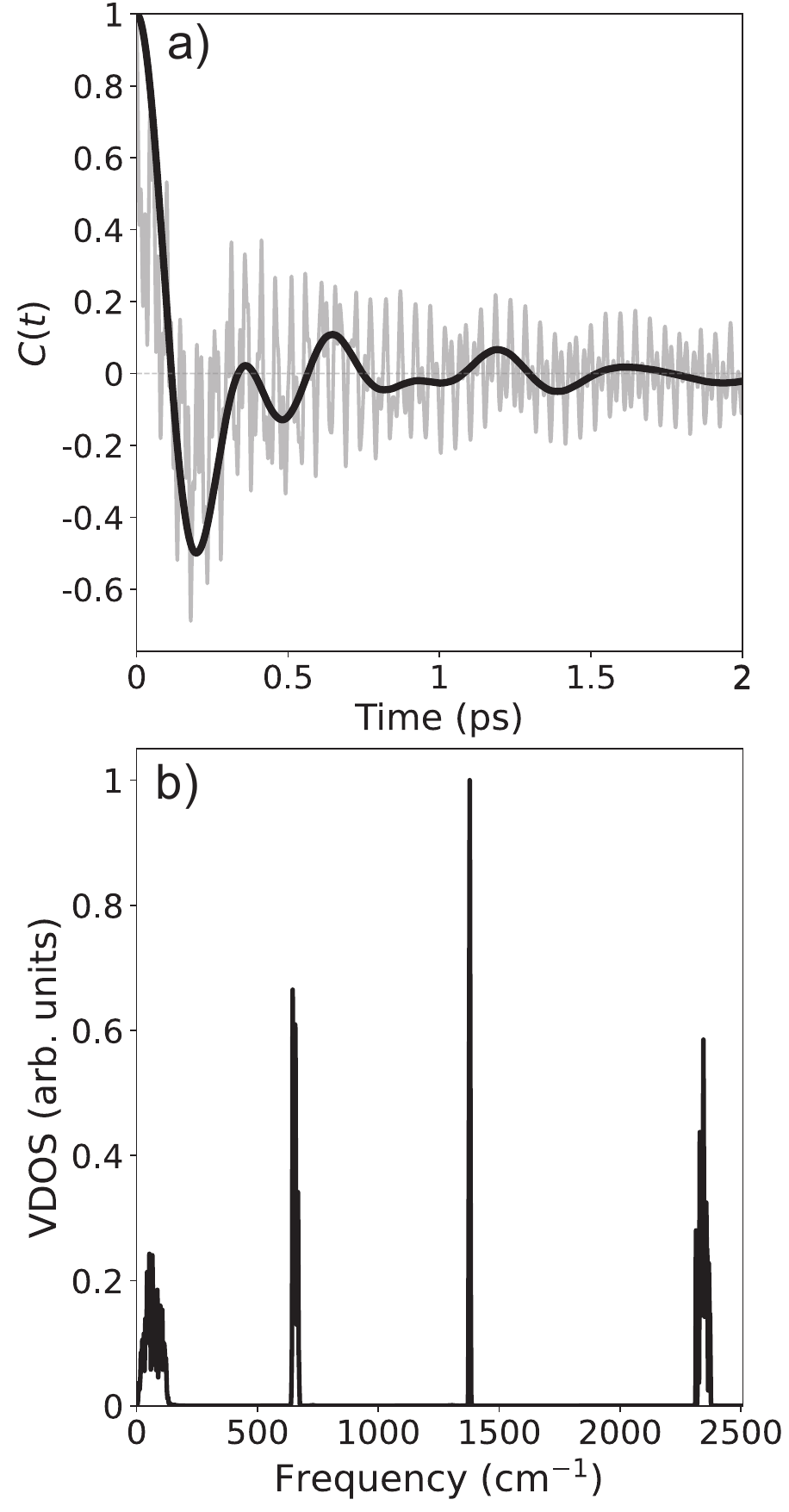}
\end{center}
\caption
{(a) Velocity autocorrelation function (VACF) and (b) vibrational density of states, $g(\omega)$, for CO$_2$ molecules in the 1\% \HH{} in CO$_2$ system. Panel (a) shows the VACF unsmoothed (gray) and smoothed over a timescale of 30~fs (black). The vibrational density of states is Gaussian smoothed with a width of 3~cm$^{-1}$. }
\label{fig:dos}
\end{figure}

Molecular-scale insights into the origins of observed slowdown of NSC with increasing \HH{} concentration can be obtained from molecular simulations,
here performed using the Orb neural network potential~\cite{Orb}. 
Motivated by the rate theory for two-phonon Raman processes, Eq.~\ref{eq:Raman_zero}, 
we computed $g(\omega)$ from velocity autocorrelation functions (VACFs). 
A representative VACF and $g(\omega)$ from a single trajectory of the 1\% \HH{} in CO$_2$ system are shown in Fig.~\ref{fig:dos}. 
For the VACF, we show both a smoothed and unsmoothed version. 
The unsmoothed VACF exhibits rapid oscillations typical of molecular vibrations. 
Smoothing over a timescale of $30$~fs filters out most of these high-frequency oscillations, leaving low-frequency features relevant to phonon modes. 
The VACF exhibits a fast decay with a pronounced global minimum near 250~fs, followed by additional oscillations until it decays to zero after nearly 2~ps. 

\FloatBarrier
%@@@@@@@@@@@@@@@@@@@@@@@@@@@@@@@@@@@@@@@@@@@@@@
%@@@@@@@@@@@@@@@@@@@@@@@@@@@@@@@@@@@@@@@@@@@@@@

%%
%

We also show $g(\omega)$ for the 1\% \HH{} system, on which a Gaussian smoothing with a width of 3~cm$^{-1}$ was performed. 
The vibrational modes of solid CO$_2$ are predicted with good accuracy: $g(\omega)$ exhibits peaks for the asymmetric stretch near 2400~cm$^{-1}$, the symmetric stretch just below 1400~cm$^{-1}$, the bending mode near 660~cm$^{-1}$, and low frequency translational/librational phonon modes below 200~cm$^{-1}$ \cite{escribano_crystallization_2013,bini_relaxation_1992}.
The good accuracy with which Orb predicts the vibrational frequencies of solid CO$_2$ is a testament to the validity of the model. 
Of particular interest here is the low-frequency region of $g(\omega)$, which is involved in the two-phonon Raman process leading to NSC. 
%

%@@@@@@@@@@@@@@@@@@@@@@@@@@@@@@@@@@@@@@@@@@@@@@
\begin{figure}[h]
\begin{center}
\includegraphics[width=0.4\textwidth]{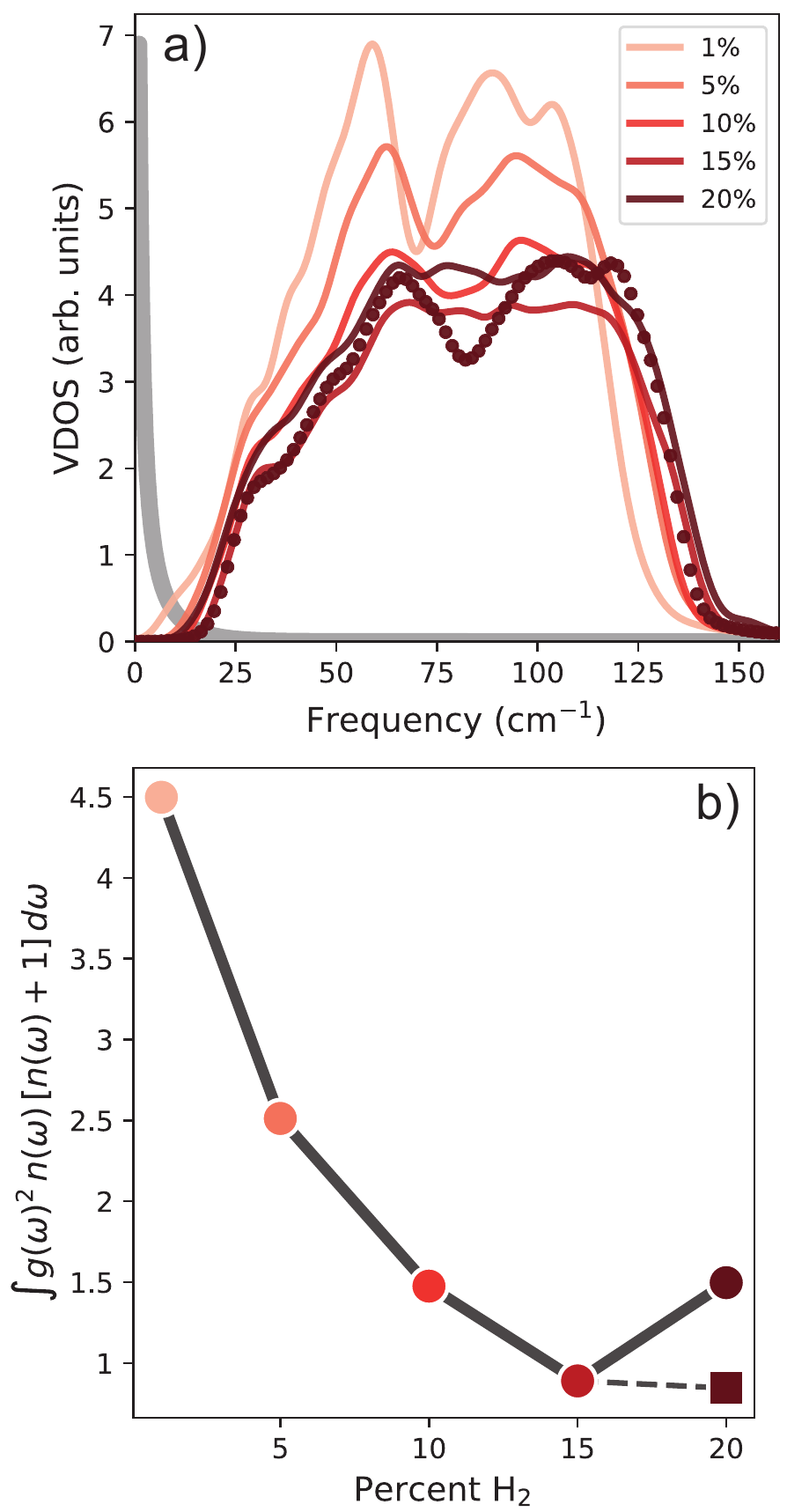}
\end{center}
\caption
{(a) CO$_2$ phonon density of states, $g(\omega)$, for all \HH{} concentrations, as well as the corresponding Bose-Einstein distribution at 10~K (gray). 
The $g(\omega)$ at each concentration is obtained as an average of many independent trajectories.
Symbols indicate $g(\omega)$ for 20\% \HH{} with a clustered arrangement. 
(b) Value of the integral contributing to the two-phonon Raman rate, equivalent to $\Gamma_{n\rightarrow n'}^{(2)}/A_{nn'}$, as a function of \HH{} concentration. 
Data point colors correspond to the corresponding curves in (a). 
Square symbol indicates the result from a clustered arrangement of \HH{} molecules.
}
\label{fig:concVDOS}
\end{figure}
%@@@@@@@@@@@@@@@@@@@@@@@@@@@@@@@@@@@@@@@@@@@@@@

%%
%
Comparison of the Bose--Einstein distribution and the low-frequency regions of $g(\omega)$ across \HH{} concentrations indicate that the NSC rate should decrease with concentration, Fig.~\ref{fig:concVDOS}.
Only the lowest frequency tail of $g(\omega)$ overlaps with the non-zero part of the Bose--Einstein distribution. 
As the concentration of \HH{} is increased, this region shifts to higher frequency and the overlap with the Bose--Einstein distribution is reduced. 
Consequently, the integral determining the rate $\Gamma_{n\rightarrow n'}^{(2)}$ decreases with increasing concentration, consistent with the experimental observations here. 
The 20\% \HH{} system is an outlier, as the rate increases upon moving from 15\% to 20\% \HH{}. 
However, as shown below, the randomly mixed or dispersed \HH{} structure used in these MD simulations is not accurate at high concentrations and using a structure typical of high concentrations restores the qualitative trend (square symbol).

\begin{figure}[h]
\begin{center}
\includegraphics[width=0.45\textwidth]{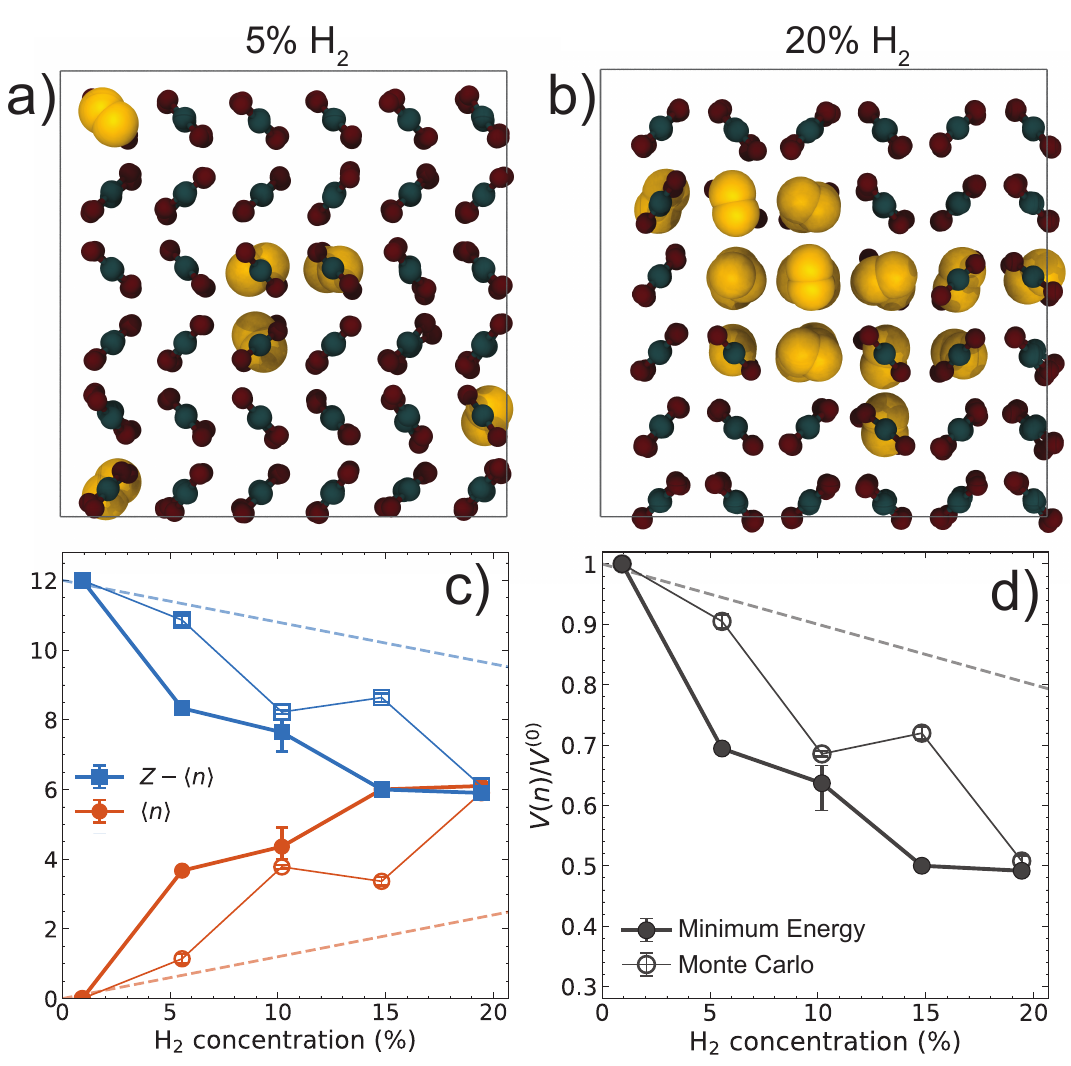}
\end{center}
\caption
{Snapshots showing typical arrangements of \HH{} molecules (yellow) in CO$_2$ at concentrations of (a) 5\% and (b) 20\% from Monte Carlo simulations. (c) Average number of \HH{} nearest neighbors, $\avg{n}$, and average number of CO$_2$ nearest neighbors, $Z-\avg{n}$, of an \HH{} molecule as a function of concentration.
(d) Crystal field on \HH{} as a function of concentration, $V(n)$, relative to that in the dilute limit, $V^{(0)}$, given by $V(n)/V^{(0)}=1-\avg{n}/Z$.
Dashed lines in (c) and (d) indicate expectations from random mixing.
Open symbols are results obtained from Monte Carlo simulations, while solid symbols are results computed from minimum energy configurations. 
Error bar on the 10\% \HH{} minimum energy configuration results is computed from several nearly degenerate configurations.
}
\label{fig:cluster}
\end{figure}

Despite the good qualitative agreement between the rates estimated with $g(\omega)$ and the experimental results for moderate concentrations, the phonon density of states does not completely explain the large decrease in NSC rate at high concentrations. 
In particular, we observe no NSC at 20\% \HH{} after 6 hours, in addition to a significantly reduced rate for 15\% \HH{}. 
We explain the dramatic decrease in NSC rate at high \HH{} concentrations through changes in the structure of the solid; \HH{} are no longer randomly dispersed. 
To quantify the propensity for \HH{} to cluster, we performed replica exchange MC simulations with moves that swap the positions of two molecules in the solid to adequately sample cluster formation (or lack thereof). 
The qualitative findings can be illustrated through the simulation snapshots in Fig.~\ref{fig:cluster}(a)--(b), which show that \HH{} molecules indeed cluster at high concentrations. 
At low concentrations, e.g. 5\% \HH{}, the \HH{} molecules do not form a single large cluster, while the \HH{} molecules in the 20\% system assemble into a single large cluster. The clustering is reminiscent of the formation dynamics of quadrupolar molecular species~\cite{buckingham_molecular_1959}.
Clustering is reflected in the average number of \HH{} molecules, $\avg{n}$, and average number of CO$_2$ molecules, $Z-\avg{n}$, in the nearest neighbor shell of a \HH{}, Fig.~\ref{fig:cluster}(c).
The coordination number is expected to follow random mixing (dashed lines) at very low \HH{} concentrations and deviates as concentration is increased, reflecting clustering of \HH{} molecules.
There, we show the coordination number averaged over configurations at 10~K from our MC simulations (open symbols), as well as from minimum energy configurations (filled symbols). 
To generate minimum energy configurations, we generated a configuration with an optimal, spherical-like \HH{} cluster and a configuration with \HH{} molecules maximally dispersed, randomized the \HH{} orientations and energy minimized those structures. 
We then computed the energies of these relaxed structures, compared with structures from the end of the MC trajectories, and used the lowest energy configuration as the minimum energy configuration; thus, they may not be true global energy minima but reflect low-energy structures of the system. 
Both the MC results and the minimum energy structures exhibit the same general trend --- $\avg{n}$ increases with \HH{} concentration, suggesting clustering of \HH{}. 
In fact, $\avg{n}$ from MC and minimum energy configurations agree quantitatively for the 20\% \HH{} system, suggesting that energetics dominates the structure at high concentrations, as discussed further below. 
In the macroscopic limit, we expect \HH{} molecules to form a single large cluster, such that $\avg{n}\rightarrow12$. 
The finite size of the simulated systems prohibits us from approaching this limit because a substantial fraction of \HH{} are at the surface of the cluster and are partially coordinated by CO$_2$ molecules; the fraction of surface molecules scales with the number of molecules in the cluster, $N$, as $N^{-1/3}$ such that convergence is slow with system size. 
%

%@@@@@@@@@@@@@@@@@@@@@@@@@@@@@@@@@@@@@@@@@@@@@@
%@@@@@@@@@@@@@@@@@@@@@@@@@@@@@@@@@@@@@@@@@@@@@@

%@@@@@@@@@@@@@@@@@@@@@@@@@@@@@@@@@@@@@@@@@@@@@@
\begin{figure*}[!t]
\centering
\includegraphics[width=0.9\textwidth]{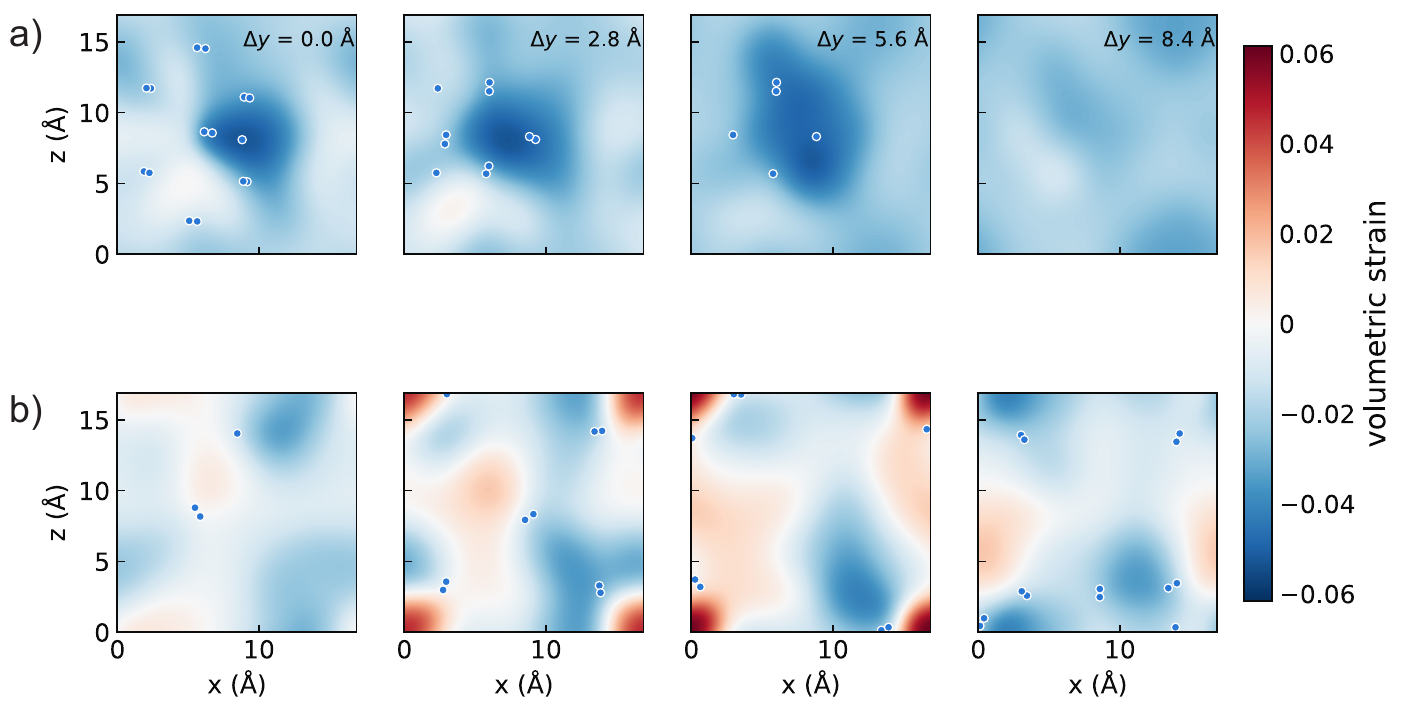}

\caption
{Strain fields for (a) clustered and (b) dispersed 20\% H$_2$ configurations shown as slices in the $xz$ plane for different distances $\Delta y$ from the center of mass of the H$_2$ cluster; the same $y$ values are used for the dispersed configurations for consistency.
Blue circles indicate positions of H atoms in the plane. Strain fields are obtained by computing the strain for each molecule and Gaussian smoothing with a width of 1.8~\AA \ to create a continuous field.}
\label{fig:strain}
\end{figure*}
%@@@@@@@@@@@@@@@@@@@@@@@@@@@@@@@@@@@@@@@@@@@@@@

%%
%
Clustering of \HH{} would non-trivially alter the prefactor $A_{nn'}$ in Eq.~\ref{eq:Raman_zero} by sequestering most of the \HH{} molecules away from the CO$_2$; only the molecules at the surface of the cluster would be in direct contact. 
Because the quadrupolar CO$_2$--\HH{} interactions are short-range, molecules in the interior of the cluster do not interact with the field from the CO$_2$. 
Clustering therefore does more than perturb the relaxation rate: it progressively removes \HH{} molecules from the host crystal field responsible for establishing the environment-induced connectivity between the ortho and para manifolds.
If we assume that the crystal field, $V$, is dominated by CO$_2$ molecules in the first coordination shell of \HH{}, we can write it as
\begin{equation}
    V = \sum_{j\in {\rm NN}}^{12} v(R_j),
\end{equation}
where NN indicates nearest neighbors, $v(R_j)$ is the interaction between \HH{} and its $j$th nearest neighbor, and there are $Z=12$ nearest neighbors in the CO$_2$ lattice. 
Upon clustering, some of the CO$_2$ neighbors of the central \HH{} are replaced by other \HH{} molecules. 
Given $n$ nearest-neighbor \HH{} molecules, we can approximate the crystal field on the central \HH{} as
\begin{equation}
    V(n) \approx \left(1-\frac{n}{Z}\right) V^{(0)},
\end{equation}
where $Z=12$ is the total number of nearest neighbors (CO$_2$ and \HH{}) and $V^{(0)}$ is the crystal field in the dilute limit where all neighbors are CO$_2$.
In the interior of a cluster, an \HH{} molecule is completely surrounded by other \HH{}, $n=12$, such that $V(n=12)=0$. 
Moreover, approximately half of the neighbors of \HH{} molecules at the cluster surface are \HH{}, reducing the crystal field by nearly half for this fraction of \HH{} molecules. 
While we do not reach the macroscopic limit in our simulations, as discussed above, we find that the crystal field is reduced beyond expectations from random mixing by clustering of \HH{} molecules, in both the MC generated and minimum energy configurations, Fig.~\ref{fig:cluster}(d). 
In this picture, increasing \HH{} concentration progressively converts the local environment from host-like to \HH{}, driving $A_{nn'}$ toward zero and closing the same conversion pathway that the host crystal field initially opened. This is consistent with the lack of NSC experimentally observed in the 20\% \HH{} system here. 
The good agreement between minimum energy configurations and MC simulations (which in principle contain entropic effects) suggest that the formation of a single large \HH{} cluster at high concentrations is driven by energetics. 
While quadrupole--quadrupole interactions exist between \HH{} molecules, they are not expected to overwhelm the combination of quadrupolar interactions with CO$_2$ and entropy of mixing that would favor the formation of dispersed or mixed structures, not clusters. 
Instead, \HH{} cluster formation emerges from minimizing the energetics of the host lattice, akin to helium bubble formation in metals~\cite{wilson_self-trapping_1981,trinkaus_energetics_1983,trinkaus_helium_2003,zhou_enabling_2021}.
Because \HH{} and CO$_2$ are not the same size and shape, substitution of a CO$_2$ molecule by \HH{} creates strain in the lattice. 
The presence of many \HH{} molecules creates multiple defect centers, each inducing an unfavorable strain field that can overlap and interact with that of others. 
The formation of a \HH{} cluster minimizes elastic energy and localizes the distortion around the cluster, whereas strain permeates the entire lattice in dispersed \HH{} configurations.
To illustrate this point, we computed the strain field for a clustered and dispersed configuration of 20\% \HH{} in CO$_2$, Fig.~\ref{fig:strain}. 
The volumetric strain or local dilatation was computed at each molecular site using established methods~\cite{falk_dynamics_1998,shimizu_theory_2007} and then a continuous field was constructed by Gaussian smoothing with a width of 1.8~\AA. 
Slices of the strain field show that the single \HH{} cluster results in compressive strain localized around the cluster; far from the cluster the strain field is smooth and nearly zero. 
In contrast, dispersed \HH{} molecules create substantial compressive and tensile strain that persists throughout the entire crystal, resulting in a significantly higher energy state than the clustered configuration. 
As a result, reduction of lattice strain can provide a driving force for \HH{} clustering. 

The strain field therefore provides the final link between \HH{} concentration and the microscopic crystal field governing NSC. Increasing \HH{} loading introduces elastic defects into the host; aggregation lowers this classical elastic energy by localizing the resulting strain, but simultaneously sequesters \HH{} from the host molecules whose crystal field enables ortho--para conversion. The concentration-dependent quantum dynamics thus emerge from a hierarchy spanning length scales: macroscopic composition determines the collective elastic response of the lattice, that response reorganizes the local molecular environment, and the local environment determines the connectivity and relaxation of the \HH{} nuclear-spin states.

%%%
\section{Implications for Environment-Defined Quantum Dynamics} \label{sec:discussion}

Nuclear-spin conversion of \HH{} exhibits an enormous range of relaxation timescales depending on the surrounding environment. 
A study of \HH{} trapped in crystalline ices of Ar, Kr, \ce{N2}, CO at 13 K reports conversion on the timescale of 2 to 3 days ($<$\SI{e-6}{\per\s}) \cite{warren_infrared_1980}, whereas studies of \HH{} encapsulated in solid \Csixty{} report relaxation rates as slow as \SI{e-12}{\per\s} \cite{yildirim_rotational_2002,mamone_rotor_2009,mamone_theory_2011}. Relaxation rates can also be quite fast, especially in non-crystalline environments: \HH{} adsorbed on amorphous solid water at 12 K relaxes at \SI{2e-4}{\per\s} \cite{hixson_experimental_1992,sugimoto_electric-field-induced_2011}, while \pHH{} and \oHH{} in amorphous solid \COO{} at 5.4 K exhibit relaxation rates of \SI{9.6e-4}{} and \SI{9.2e-4}{\per\s}, respectively \cite{yamakawa_fast_2020}. The comparatively rapid conversion observed here in crystalline \COO{} and \NNO{} further illustrates the sensitivity of nuclear-spin quantum dynamics to the surrounding environment, even in the absence of a paramagnet \cite{nielsen_paramagnetic_1967,wigner_uber_1933,ilisca_electron_2014}.

Our experiments demonstrate that both the identity and composition of the surrounding medium influence nuclear-spin relaxation. At low \HH{} loading, conversion in crystalline \COO{} is approximately three times slower than in crystalline \NNO{}, despite the two solids sharing the same crystal symmetry under our experimental conditions. The main difference between these two crystals is the presence of a permanent dipole moment in \NNO{} that contributes to dipole--quadrupole interactions absent in \COO{} and provides additional pathways for nuclear-spin conversion \cite{mclane_environment-imposed_2026}. However, this host dependence diminishes as \HH{} loading increases, with comparable relaxation rates observed between the two crystals at 10\% \HH{}. This observation is consistent with a picture of a progressively \HH{}-rich local environment that reduces the influence of the host crystal field. The present work tests this observation in simulations of the \COO{} crystal, and a corresponding mechanism likely exists in the \NNO{} crystal. However, the longer-ranged dipole--quadrupole interaction ($\sim r^{-4}$) in \NNO{} may require more complete replacement of the surrounding coordination shells than the quadrupole--quadrupole interaction ($\sim r^{-5}$) in \COO{}.

The combined experimental and computational results provide a microscopic description of how the surrounding environment influences the dynamics of an effective molecular two-state system. In our previous work, we demonstrated that confinement selectively enables nuclear-spin conversion between the ortho ($j=1,m=0$) and para ($j=0$) states of \HH{}, providing an environment-defined pathway between states that are otherwise disconnected in the isolated molecule. Here, the agreement between the measured ortho-loss and para-growth rates, together with the absence of observable relaxation from the $|m|=1$ ortho states, is consistent with this state-selective picture. More importantly, we find that the dynamics of this conversion pathway are strongly dependent on the surrounding material. Within a two-phonon Raman description, increasing \HH{} loading modifies both the vibrational spectrum of the host and the strength of the local crystal-field coupling, progressively suppressing nuclear-spin conversion.

This system therefore provides a particularly useful experimental realization of an open quantum system in which the microscopic quantum states are suitably characterized, but their dynamics depend sensitively on interactions with the surrounding environment. The classical elastic response of the host introduces an additional layer of complexity: changes in molecular composition generate strain that reorganizes the local environment experienced by \HH{}, thereby modifying its nuclear-spin relaxation dynamics. The resulting connection between collective material behavior and microscopic quantum-state evolution illustrates how environmental structure and dynamics can influence the lifetime of an effective molecular two-state subsystem. More broadly, these findings provide a framework for understanding the role of condensed-phase environments in molecular quantum-state preparation and preservation.

\begin{acknowledgments}

This material is based upon work supported by the U.S. Department of Energy, Office of Science Early Career Research Program, Office of Basic Energy Sciences under Award Number DE-SC0024262. NM was partially supported by NSF award DMS-2231533.  RCR acknowledges the Office of Advanced Research Computing (OARC) at Rutgers, The State University of New Jersey for providing access to the Amarel cluster and associated research computing resources that have contributed to the results reported here.
\end{acknowledgments}
\section*{Data Availability}
\label{sec:data_availability}
The data that support the findings of this article are openly available \cite{data_availability}.

\nocite{orbv3,omol25,wb97mv,vv10,def2,def2d,rijcosx,def2j,orca2012,orca6}

%\bibliographystyle{apsrev4-2} % use for Phys Rev journals
%\bibliography{Dodson_lab_UMD_total}% for journal submission

\begin{thebibliography}{72}%
\makeatletter
\providecommand \@ifxundefined [1]{%
 \@ifx{#1\undefined}
}%
\providecommand \@ifnum [1]{%
 \ifnum #1\expandafter \@firstoftwo
 \else \expandafter \@secondoftwo
 \fi
}%
\providecommand \@ifx [1]{%
 \ifx #1\expandafter \@firstoftwo
 \else \expandafter \@secondoftwo
 \fi
}%
\providecommand \natexlab [1]{#1}%
\providecommand \enquote  [1]{``#1''}%
\providecommand \bibnamefont  [1]{#1}%
\providecommand \bibfnamefont [1]{#1}%
\providecommand \citenamefont [1]{#1}%
\providecommand \href@noop [0]{\@secondoftwo}%
\providecommand \href [0]{\begingroup \@sanitize@url \@href}%
\providecommand \@href[1]{\@@startlink{#1}\@@href}%
\providecommand \@@href[1]{\endgroup#1\@@endlink}%
\providecommand \@sanitize@url [0]{\catcode `\\12\catcode `\$12\catcode `\&12\catcode `\#12\catcode `\^12\catcode `\_12\catcode `\%12\relax}%
\providecommand \@@startlink[1]{}%
\providecommand \@@endlink[0]{}%
\providecommand \url  [0]{\begingroup\@sanitize@url \@url }%
\providecommand \@url [1]{\endgroup\@href {#1}{\urlprefix }}%
\providecommand \urlprefix  [0]{URL }%
\providecommand \Eprint [0]{\href }%
\providecommand \doibase [0]{https://doi.org/}%
\providecommand \selectlanguage [0]{\@gobble}%
\providecommand \bibinfo  [0]{\@secondoftwo}%
\providecommand \bibfield  [0]{\@secondoftwo}%
\providecommand \translation [1]{[#1]}%
\providecommand \BibitemOpen [0]{}%
\providecommand \bibitemStop [0]{}%
\providecommand \bibitemNoStop [0]{.\EOS\space}%
\providecommand \EOS [0]{\spacefactor3000\relax}%
\providecommand \BibitemShut  [1]{\csname bibitem#1\endcsname}%
\let\auto@bib@innerbib\@empty
%</preamble>
\bibitem [{\citenamefont {Mattes}\ \emph {et~al.}(2025)\citenamefont {Mattes}, \citenamefont {Cabot}, \citenamefont {Carollo},\ and\ \citenamefont {Lesanovsky}}]{mattes_designing_2025}%
  \BibitemOpen
  \bibfield  {author} {\bibinfo {author} {\bibfnamefont {R.}~\bibnamefont {Mattes}}, \bibinfo {author} {\bibfnamefont {A.}~\bibnamefont {Cabot}}, \bibinfo {author} {\bibfnamefont {F.}~\bibnamefont {Carollo}},\ and\ \bibinfo {author} {\bibfnamefont {I.}~\bibnamefont {Lesanovsky}},\ }\href {https://doi.org/10.1103/5gh9-nmv8} {\bibfield  {journal} {\bibinfo  {journal} {Physical Review Letters}\ }\textbf {\bibinfo {volume} {135}},\ \bibinfo {pages} {230402} (\bibinfo {year} {2025})}\BibitemShut {NoStop}%
\bibitem [{\citenamefont {Ilias}\ \emph {et~al.}(2022)\citenamefont {Ilias}, \citenamefont {Yang}, \citenamefont {Huelga},\ and\ \citenamefont {Plenio}}]{ilias_criticality-enhanced_2022}%
  \BibitemOpen
  \bibfield  {author} {\bibinfo {author} {\bibfnamefont {T.}~\bibnamefont {Ilias}}, \bibinfo {author} {\bibfnamefont {D.}~\bibnamefont {Yang}}, \bibinfo {author} {\bibfnamefont {S.~F.}\ \bibnamefont {Huelga}},\ and\ \bibinfo {author} {\bibfnamefont {M.~B.}\ \bibnamefont {Plenio}},\ }\href {https://doi.org/10.1103/PRXQuantum.3.010354} {\bibfield  {journal} {\bibinfo  {journal} {PRX Quantum}\ }\textbf {\bibinfo {volume} {3}},\ \bibinfo {pages} {010354} (\bibinfo {year} {2022})}\BibitemShut {NoStop}%
\bibitem [{\citenamefont {Bermudez}\ \emph {et~al.}(2013)\citenamefont {Bermudez}, \citenamefont {Schaetz},\ and\ \citenamefont {Plenio}}]{bermudez_dissipation-assisted_2013}%
  \BibitemOpen
  \bibfield  {author} {\bibinfo {author} {\bibfnamefont {A.}~\bibnamefont {Bermudez}}, \bibinfo {author} {\bibfnamefont {T.}~\bibnamefont {Schaetz}},\ and\ \bibinfo {author} {\bibfnamefont {M.~B.}\ \bibnamefont {Plenio}},\ }\href {https://doi.org/10.1103/PhysRevLett.110.110502} {\bibfield  {journal} {\bibinfo  {journal} {Physical Review Letters}\ }\textbf {\bibinfo {volume} {110}},\ \bibinfo {pages} {110502} (\bibinfo {year} {2013})}\BibitemShut {NoStop}%
\bibitem [{\citenamefont {Fallas~Padilla}\ \emph {et~al.}(2025)\citenamefont {Fallas~Padilla}, \citenamefont {So}, \citenamefont {Menon}, \citenamefont {Zhuravel}, \citenamefont {Pu},\ and\ \citenamefont {Pagano}}]{fallas_padilla_delocalized_2025}%
  \BibitemOpen
  \bibfield  {author} {\bibinfo {author} {\bibfnamefont {D.}~\bibnamefont {Fallas~Padilla}}, \bibinfo {author} {\bibfnamefont {V.}~\bibnamefont {So}}, \bibinfo {author} {\bibfnamefont {A.}~\bibnamefont {Menon}}, \bibinfo {author} {\bibfnamefont {R.}~\bibnamefont {Zhuravel}}, \bibinfo {author} {\bibfnamefont {H.}~\bibnamefont {Pu}},\ and\ \bibinfo {author} {\bibfnamefont {G.}~\bibnamefont {Pagano}},\ }\href {https://doi.org/10.1103/bxwl-sbsn} {\bibfield  {journal} {\bibinfo  {journal} {PRX Quantum}\ }\textbf {\bibinfo {volume} {6}},\ \bibinfo {pages} {040301} (\bibinfo {year} {2025})}\BibitemShut {NoStop}%
\bibitem [{\citenamefont {Hartmann}\ and\ \citenamefont {Carleo}(2019)}]{hartmann_neural-network_2019}%
  \BibitemOpen
  \bibfield  {author} {\bibinfo {author} {\bibfnamefont {M.~J.}\ \bibnamefont {Hartmann}}\ and\ \bibinfo {author} {\bibfnamefont {G.}~\bibnamefont {Carleo}},\ }\href {https://doi.org/10.1103/PhysRevLett.122.250502} {\bibfield  {journal} {\bibinfo  {journal} {Physical Review Letters}\ }\textbf {\bibinfo {volume} {122}},\ \bibinfo {pages} {250502} (\bibinfo {year} {2019})}\BibitemShut {NoStop}%
\bibitem [{\citenamefont {Rotter}\ and\ \citenamefont {Bird}(2015)}]{rotter_review_2015}%
  \BibitemOpen
  \bibfield  {author} {\bibinfo {author} {\bibfnamefont {I.}~\bibnamefont {Rotter}}\ and\ \bibinfo {author} {\bibfnamefont {J.~P.}\ \bibnamefont {Bird}},\ }\href {https://doi.org/10.1088/0034-4885/78/11/114001} {\bibfield  {journal} {\bibinfo  {journal} {Reports on Progress in Physics}\ }\textbf {\bibinfo {volume} {78}},\ \bibinfo {pages} {114001} (\bibinfo {year} {2015})}\BibitemShut {NoStop}%
\bibitem [{\citenamefont {Kapral}(2015)}]{kapral_quantum_2015}%
  \BibitemOpen
  \bibfield  {author} {\bibinfo {author} {\bibfnamefont {R.}~\bibnamefont {Kapral}},\ }\href {https://doi.org/10.1088/0953-8984/27/7/073201} {\bibfield  {journal} {\bibinfo  {journal} {Journal of Physics: Condensed Matter}\ }\textbf {\bibinfo {volume} {27}},\ \bibinfo {pages} {073201} (\bibinfo {year} {2015})}\BibitemShut {NoStop}%
\bibitem [{\citenamefont {Bader}(2009)}]{bader_confined_2009}%
  \BibitemOpen
  \bibfield  {author} {\bibinfo {author} {\bibfnamefont {R.~F.}\ \bibnamefont {Bader}},\ }in\ \href {https://doi.org/10.1016/S0065-3276(09)00613-3} {\emph {\bibinfo {booktitle} {Advances in {Quantum} {Chemistry}}}},\ Vol.~\bibinfo {volume} {57}\ (\bibinfo  {publisher} {Elsevier},\ \bibinfo {year} {2009})\ pp.\ \bibinfo {pages} {285--318}\BibitemShut {NoStop}%
\bibitem [{\citenamefont {Bayliss}\ \emph {et~al.}(2022)\citenamefont {Bayliss}, \citenamefont {Deb}, \citenamefont {Laorenza}, \citenamefont {Onizhuk}, \citenamefont {Galli}, \citenamefont {Freedman},\ and\ \citenamefont {Awschalom}}]{bayliss_enhancing_2022}%
  \BibitemOpen
  \bibfield  {author} {\bibinfo {author} {\bibfnamefont {S.}~\bibnamefont {Bayliss}}, \bibinfo {author} {\bibfnamefont {P.}~\bibnamefont {Deb}}, \bibinfo {author} {\bibfnamefont {D.}~\bibnamefont {Laorenza}}, \bibinfo {author} {\bibfnamefont {M.}~\bibnamefont {Onizhuk}}, \bibinfo {author} {\bibfnamefont {G.}~\bibnamefont {Galli}}, \bibinfo {author} {\bibfnamefont {D.}~\bibnamefont {Freedman}},\ and\ \bibinfo {author} {\bibfnamefont {D.}~\bibnamefont {Awschalom}},\ }\href {https://doi.org/10.1103/PhysRevX.12.031028} {\bibfield  {journal} {\bibinfo  {journal} {Physical Review X}\ }\textbf {\bibinfo {volume} {12}},\ \bibinfo {pages} {031028} (\bibinfo {year} {2022})}\BibitemShut {NoStop}%
\bibitem [{\citenamefont {Albert}\ \emph {et~al.}(2020)\citenamefont {Albert}, \citenamefont {Covey},\ and\ \citenamefont {Preskill}}]{albert_robust_2020}%
  \BibitemOpen
  \bibfield  {author} {\bibinfo {author} {\bibfnamefont {V.~V.}\ \bibnamefont {Albert}}, \bibinfo {author} {\bibfnamefont {J.~P.}\ \bibnamefont {Covey}},\ and\ \bibinfo {author} {\bibfnamefont {J.}~\bibnamefont {Preskill}},\ }\href {https://doi.org/10.1103/PhysRevX.10.031050} {\bibfield  {journal} {\bibinfo  {journal} {Physical Review X}\ }\textbf {\bibinfo {volume} {10}},\ \bibinfo {pages} {031050} (\bibinfo {year} {2020})}\BibitemShut {NoStop}%
\bibitem [{\citenamefont {Leggett}\ \emph {et~al.}(1987)\citenamefont {Leggett}, \citenamefont {Chakravarty}, \citenamefont {Dorsey}, \citenamefont {Fisher}, \citenamefont {Garg},\ and\ \citenamefont {Zwerger}}]{leggett_dynamics_1987}%
  \BibitemOpen
  \bibfield  {author} {\bibinfo {author} {\bibfnamefont {A.~J.}\ \bibnamefont {Leggett}}, \bibinfo {author} {\bibfnamefont {S.}~\bibnamefont {Chakravarty}}, \bibinfo {author} {\bibfnamefont {A.~T.}\ \bibnamefont {Dorsey}}, \bibinfo {author} {\bibfnamefont {M.~P.~A.}\ \bibnamefont {Fisher}}, \bibinfo {author} {\bibfnamefont {A.}~\bibnamefont {Garg}},\ and\ \bibinfo {author} {\bibfnamefont {W.}~\bibnamefont {Zwerger}},\ }\href {https://doi.org/10.1103/RevModPhys.59.1} {\bibfield  {journal} {\bibinfo  {journal} {Reviews of Modern Physics}\ }\textbf {\bibinfo {volume} {59}},\ \bibinfo {pages} {1} (\bibinfo {year} {1987})}\BibitemShut {NoStop}%
\bibitem [{\citenamefont {Wigner}(1933)}]{wigner_uber_1933}%
  \BibitemOpen
  \bibfield  {author} {\bibinfo {author} {\bibfnamefont {E.}~\bibnamefont {Wigner}},\ }\href@noop {} {\bibfield  {journal} {\bibinfo  {journal} {Z. Phys. Chem. B}\ }\textbf {\bibinfo {volume} {23}},\ \bibinfo {pages} {28} (\bibinfo {year} {1933})}\BibitemShut {NoStop}%
\bibitem [{\citenamefont {Farkas}(1935)}]{farkas_orthohydrogen_1935}%
  \BibitemOpen
  \bibfield  {author} {\bibinfo {author} {\bibfnamefont {A.}~\bibnamefont {Farkas}},\ }\href@noop {} {\emph {\bibinfo {title} {Orthohydrogen, {Parahydrogen} and {Heavy} {Hydrogen}}}},\ Cambridge {Series} of {Physical} {Chemistry}\ (\bibinfo  {publisher} {Cambridge University Press},\ \bibinfo {address} {Cambridge},\ \bibinfo {year} {1935})\BibitemShut {NoStop}%
\bibitem [{\citenamefont {Van~Kranendonk}(1958)}]{van_kranendonk_induced_1958}%
  \BibitemOpen
  \bibfield  {author} {\bibinfo {author} {\bibfnamefont {J.}~\bibnamefont {Van~Kranendonk}},\ }\href {https://doi.org/10.1016/S0031-8914(58)95223-6} {\bibfield  {journal} {\bibinfo  {journal} {Physica}\ }\textbf {\bibinfo {volume} {24}},\ \bibinfo {pages} {347} (\bibinfo {year} {1958})}\BibitemShut {NoStop}%
\bibitem [{\citenamefont {Yildirim}\ and\ \citenamefont {Harris}(2002)}]{yildirim_rotational_2002}%
  \BibitemOpen
  \bibfield  {author} {\bibinfo {author} {\bibfnamefont {T.}~\bibnamefont {Yildirim}}\ and\ \bibinfo {author} {\bibfnamefont {A.~B.}\ \bibnamefont {Harris}},\ }\href {https://doi.org/10.1103/PhysRevB.66.214301} {\bibfield  {journal} {\bibinfo  {journal} {Physical Review B}\ }\textbf {\bibinfo {volume} {66}},\ \bibinfo {pages} {214301} (\bibinfo {year} {2002})}\BibitemShut {NoStop}%
\bibitem [{\citenamefont {Kohama}\ \emph {et~al.}(2009)\citenamefont {Kohama}, \citenamefont {Rachi}, \citenamefont {Jing}, \citenamefont {Li}, \citenamefont {Tang}, \citenamefont {Kumashiro}, \citenamefont {Izumisawa}, \citenamefont {Kawaji}, \citenamefont {Atake}, \citenamefont {Sawa}, \citenamefont {Murata}, \citenamefont {Komatsu},\ and\ \citenamefont {Tanigaki}}]{kohama_rotational_2009}%
  \BibitemOpen
  \bibfield  {author} {\bibinfo {author} {\bibfnamefont {Y.}~\bibnamefont {Kohama}}, \bibinfo {author} {\bibfnamefont {T.}~\bibnamefont {Rachi}}, \bibinfo {author} {\bibfnamefont {J.}~\bibnamefont {Jing}}, \bibinfo {author} {\bibfnamefont {Z.}~\bibnamefont {Li}}, \bibinfo {author} {\bibfnamefont {J.}~\bibnamefont {Tang}}, \bibinfo {author} {\bibfnamefont {R.}~\bibnamefont {Kumashiro}}, \bibinfo {author} {\bibfnamefont {S.}~\bibnamefont {Izumisawa}}, \bibinfo {author} {\bibfnamefont {H.}~\bibnamefont {Kawaji}}, \bibinfo {author} {\bibfnamefont {T.}~\bibnamefont {Atake}}, \bibinfo {author} {\bibfnamefont {H.}~\bibnamefont {Sawa}}, \bibinfo {author} {\bibfnamefont {Y.}~\bibnamefont {Murata}}, \bibinfo {author} {\bibfnamefont {K.}~\bibnamefont {Komatsu}},\ and\ \bibinfo {author} {\bibfnamefont {K.}~\bibnamefont {Tanigaki}},\ }\href {https://doi.org/10.1103/PhysRevLett.103.073001} {\bibfield  {journal} {\bibinfo  {journal} {Physical Review Letters}\ }\textbf {\bibinfo {volume} {103}},\ \bibinfo {pages} {073001}
  (\bibinfo {year} {2009})}\BibitemShut {NoStop}%
\bibitem [{\citenamefont {Ilisca}\ and\ \citenamefont {Ghiglieno}(2014)}]{ilisca_electron_2014}%
  \BibitemOpen
  \bibfield  {author} {\bibinfo {author} {\bibfnamefont {E.}~\bibnamefont {Ilisca}}\ and\ \bibinfo {author} {\bibfnamefont {F.}~\bibnamefont {Ghiglieno}},\ }\href {https://doi.org/10.1140/epjb/e2014-50282-2} {\bibfield  {journal} {\bibinfo  {journal} {The European Physical Journal B}\ }\textbf {\bibinfo {volume} {87}},\ \bibinfo {pages} {235} (\bibinfo {year} {2014})}\BibitemShut {NoStop}%
\bibitem [{\citenamefont {Herzberg}\ and\ \citenamefont {Crawford}(1946)}]{herzberg_infrared_1946}%
  \BibitemOpen
  \bibfield  {author} {\bibinfo {author} {\bibfnamefont {G.}~\bibnamefont {Herzberg}}\ and\ \bibinfo {author} {\bibfnamefont {B.~L.}\ \bibnamefont {Crawford}},\ }\href {https://doi.org/10.1021/j150447a021} {\bibfield  {journal} {\bibinfo  {journal} {The Journal of Physical Chemistry}\ }\textbf {\bibinfo {volume} {50}},\ \bibinfo {pages} {288} (\bibinfo {year} {1946})}\BibitemShut {NoStop}%
\bibitem [{\citenamefont {Herzberg}(1950)}]{herzberg_molecular_1950}%
  \BibitemOpen
  \bibfield  {author} {\bibinfo {author} {\bibfnamefont {G.}~\bibnamefont {Herzberg}},\ }\href@noop {} {\emph {\bibinfo {title} {Molecular {Spectra} and {Molecular} {Structure}. {I}. {Spectra} of {Diatomic} {Molecules}}}}\ (\bibinfo  {publisher} {D. Van Nostrand Company, Inc},\ \bibinfo {address} {Princeton, NJ},\ \bibinfo {year} {1950})\BibitemShut {NoStop}%
\bibitem [{\citenamefont {Silvera}(1980)}]{silvera_solid_1980}%
  \BibitemOpen
  \bibfield  {author} {\bibinfo {author} {\bibfnamefont {I.~F.}\ \bibnamefont {Silvera}},\ }\href {https://doi.org/10.1103/RevModPhys.52.393} {\bibfield  {journal} {\bibinfo  {journal} {Reviews of Modern Physics}\ }\textbf {\bibinfo {volume} {52}},\ \bibinfo {pages} {393} (\bibinfo {year} {1980})}\BibitemShut {NoStop}%
\bibitem [{\citenamefont {Curl}\ \emph {et~al.}(1967)\citenamefont {Curl}, \citenamefont {Kasper},\ and\ \citenamefont {Pitzer}}]{curl_nuclear_1967}%
  \BibitemOpen
  \bibfield  {author} {\bibinfo {author} {\bibfnamefont {R.~F.}\ \bibnamefont {Curl}}, \bibinfo {author} {\bibfnamefont {J.~V.~V.}\ \bibnamefont {Kasper}},\ and\ \bibinfo {author} {\bibfnamefont {K.~S.}\ \bibnamefont {Pitzer}},\ }\href {https://doi.org/10.1063/1.1841193} {\bibfield  {journal} {\bibinfo  {journal} {The Journal of Chemical Physics}\ }\textbf {\bibinfo {volume} {46}},\ \bibinfo {pages} {3220} (\bibinfo {year} {1967})}\BibitemShut {NoStop}%
\bibitem [{\citenamefont {Nielsen}\ and\ \citenamefont {Dahler}(1967)}]{nielsen_paramagnetic_1967}%
  \BibitemOpen
  \bibfield  {author} {\bibinfo {author} {\bibfnamefont {S.~E.}\ \bibnamefont {Nielsen}}\ and\ \bibinfo {author} {\bibfnamefont {J.~S.}\ \bibnamefont {Dahler}},\ }\href {https://doi.org/10.1063/1.1840734} {\bibfield  {journal} {\bibinfo  {journal} {The Journal of Chemical Physics}\ }\textbf {\bibinfo {volume} {46}},\ \bibinfo {pages} {732} (\bibinfo {year} {1967})}\BibitemShut {NoStop}%
\bibitem [{\citenamefont {Schweizer}\ \emph {et~al.}(1979)\citenamefont {Schweizer}, \citenamefont {Washburn},\ and\ \citenamefont {Meyer}}]{schweizer_nmr_1979}%
  \BibitemOpen
  \bibfield  {author} {\bibinfo {author} {\bibfnamefont {R.}~\bibnamefont {Schweizer}}, \bibinfo {author} {\bibfnamefont {S.}~\bibnamefont {Washburn}},\ and\ \bibinfo {author} {\bibfnamefont {H.}~\bibnamefont {Meyer}},\ }\href {https://doi.org/10.1007/BF00119191} {\bibfield  {journal} {\bibinfo  {journal} {Journal of Low Temperature Physics}\ }\textbf {\bibinfo {volume} {37}},\ \bibinfo {pages} {289} (\bibinfo {year} {1979})}\BibitemShut {NoStop}%
\bibitem [{\citenamefont {FitzGerald}\ \emph {et~al.}(2008)\citenamefont {FitzGerald}, \citenamefont {Allen}, \citenamefont {Landerman}, \citenamefont {Hopkins}, \citenamefont {Matters}, \citenamefont {Myers},\ and\ \citenamefont {Rowsell}}]{fitzgerald_quantum_2008}%
  \BibitemOpen
  \bibfield  {author} {\bibinfo {author} {\bibfnamefont {S.~A.}\ \bibnamefont {FitzGerald}}, \bibinfo {author} {\bibfnamefont {K.}~\bibnamefont {Allen}}, \bibinfo {author} {\bibfnamefont {P.}~\bibnamefont {Landerman}}, \bibinfo {author} {\bibfnamefont {J.}~\bibnamefont {Hopkins}}, \bibinfo {author} {\bibfnamefont {J.}~\bibnamefont {Matters}}, \bibinfo {author} {\bibfnamefont {R.}~\bibnamefont {Myers}},\ and\ \bibinfo {author} {\bibfnamefont {J.~L.~C.}\ \bibnamefont {Rowsell}},\ }\href {https://doi.org/10.1103/PhysRevB.77.224301} {\bibfield  {journal} {\bibinfo  {journal} {Physical Review B}\ }\textbf {\bibinfo {volume} {77}},\ \bibinfo {pages} {224301} (\bibinfo {year} {2008})}\BibitemShut {NoStop}%
\bibitem [{\citenamefont {Di~Cataldo}\ \emph {et~al.}(2024)\citenamefont {Di~Cataldo}, \citenamefont {Rescigno}, \citenamefont {Monacelli}, \citenamefont {Ranieri}, \citenamefont {Gaal}, \citenamefont {Klotz}, \citenamefont {Ollivier}, \citenamefont {Koza}, \citenamefont {De~Michele},\ and\ \citenamefont {Bove}}]{di_cataldo_giant_2024}%
  \BibitemOpen
  \bibfield  {author} {\bibinfo {author} {\bibfnamefont {S.}~\bibnamefont {Di~Cataldo}}, \bibinfo {author} {\bibfnamefont {M.}~\bibnamefont {Rescigno}}, \bibinfo {author} {\bibfnamefont {L.}~\bibnamefont {Monacelli}}, \bibinfo {author} {\bibfnamefont {U.}~\bibnamefont {Ranieri}}, \bibinfo {author} {\bibfnamefont {R.}~\bibnamefont {Gaal}}, \bibinfo {author} {\bibfnamefont {S.}~\bibnamefont {Klotz}}, \bibinfo {author} {\bibfnamefont {J.}~\bibnamefont {Ollivier}}, \bibinfo {author} {\bibfnamefont {M.~M.}\ \bibnamefont {Koza}}, \bibinfo {author} {\bibfnamefont {C.}~\bibnamefont {De~Michele}},\ and\ \bibinfo {author} {\bibfnamefont {L.~E.}\ \bibnamefont {Bove}},\ }\href {https://doi.org/10.1103/PhysRevLett.133.236101} {\bibfield  {journal} {\bibinfo  {journal} {Physical Review Letters}\ }\textbf {\bibinfo {volume} {133}},\ \bibinfo {pages} {236101} (\bibinfo {year} {2024})}\BibitemShut {NoStop}%
\bibitem [{\citenamefont {McLane}\ \emph {et~al.}(2026)\citenamefont {McLane}, \citenamefont {Duckett},\ and\ \citenamefont {Dodson}}]{mclane_environment-imposed_2026}%
  \BibitemOpen
  \bibfield  {author} {\bibinfo {author} {\bibfnamefont {N.}~\bibnamefont {McLane}}, \bibinfo {author} {\bibfnamefont {L.}~\bibnamefont {Duckett}},\ and\ \bibinfo {author} {\bibfnamefont {L.~G.}\ \bibnamefont {Dodson}},\ }\href {https://doi.org/10.1103/2yw9-7h62} {\bibfield  {journal} {\bibinfo  {journal} {Physical Review Letters}\ }\textbf {\bibinfo {volume} {136}},\ \bibinfo {pages} {178002} (\bibinfo {year} {2026})}\BibitemShut {NoStop}%
\bibitem [{\citenamefont {Chisholm}\ and\ \citenamefont {Welsh}(1954)}]{chisholm_induced_1954}%
  \BibitemOpen
  \bibfield  {author} {\bibinfo {author} {\bibfnamefont {D.~A.}\ \bibnamefont {Chisholm}}\ and\ \bibinfo {author} {\bibfnamefont {H.~L.}\ \bibnamefont {Welsh}},\ }\href {https://doi.org/10.1139/p54-028} {\bibfield  {journal} {\bibinfo  {journal} {Canadian Journal of Physics}\ }\textbf {\bibinfo {volume} {32}},\ \bibinfo {pages} {291} (\bibinfo {year} {1954})}\BibitemShut {NoStop}%
\bibitem [{\citenamefont {Hockey}\ \emph {et~al.}(2022)\citenamefont {Hockey}, \citenamefont {Vlahos}, \citenamefont {Howard}, \citenamefont {Palko},\ and\ \citenamefont {Dodson}}]{hockey_weakly_2022}%
  \BibitemOpen
  \bibfield  {author} {\bibinfo {author} {\bibfnamefont {E.~K.}\ \bibnamefont {Hockey}}, \bibinfo {author} {\bibfnamefont {K.}~\bibnamefont {Vlahos}}, \bibinfo {author} {\bibfnamefont {T.}~\bibnamefont {Howard}}, \bibinfo {author} {\bibfnamefont {J.}~\bibnamefont {Palko}},\ and\ \bibinfo {author} {\bibfnamefont {L.~G.}\ \bibnamefont {Dodson}},\ }\href {https://doi.org/10.1021/acs.jpca.2c00716} {\bibfield  {journal} {\bibinfo  {journal} {The Journal of Physical Chemistry A}\ }\textbf {\bibinfo {volume} {126}},\ \bibinfo {pages} {3110} (\bibinfo {year} {2022})}\BibitemShut {NoStop}%
\bibitem [{\citenamefont {Hockey}\ \emph {et~al.}(2024)\citenamefont {Hockey}, \citenamefont {McLane}, \citenamefont {Vlahos}, \citenamefont {McCaslin},\ and\ \citenamefont {Dodson}}]{hockey_matrix-formation_2024}%
  \BibitemOpen
  \bibfield  {author} {\bibinfo {author} {\bibfnamefont {E.~K.}\ \bibnamefont {Hockey}}, \bibinfo {author} {\bibfnamefont {N.}~\bibnamefont {McLane}}, \bibinfo {author} {\bibfnamefont {K.}~\bibnamefont {Vlahos}}, \bibinfo {author} {\bibfnamefont {L.~M.}\ \bibnamefont {McCaslin}},\ and\ \bibinfo {author} {\bibfnamefont {L.~G.}\ \bibnamefont {Dodson}},\ }\bibfield  {journal} {\bibinfo  {journal} {Journal of Chemical Physics}\ }\textbf {\bibinfo {volume} {160}},\ \href {https://doi.org/10.1063/5.0188433} {10.1063/5.0188433} (\bibinfo {year} {2024})\BibitemShut {NoStop}%
\bibitem [{\citenamefont {Miyamoto}\ \emph {et~al.}(2008)\citenamefont {Miyamoto}, \citenamefont {Fushitani}, \citenamefont {Ando},\ and\ \citenamefont {Momose}}]{miyamoto_nuclear_2008}%
  \BibitemOpen
  \bibfield  {author} {\bibinfo {author} {\bibfnamefont {Y.}~\bibnamefont {Miyamoto}}, \bibinfo {author} {\bibfnamefont {M.}~\bibnamefont {Fushitani}}, \bibinfo {author} {\bibfnamefont {D.}~\bibnamefont {Ando}},\ and\ \bibinfo {author} {\bibfnamefont {T.}~\bibnamefont {Momose}},\ }\href {https://doi.org/10.1063/1.2889002} {\bibfield  {journal} {\bibinfo  {journal} {The Journal of Chemical Physics}\ }\textbf {\bibinfo {volume} {128}},\ \bibinfo {pages} {114502} (\bibinfo {year} {2008})}\BibitemShut {NoStop}%
\bibitem [{\citenamefont {Fedders}(1979)}]{fedders_nuclear-spin-lattice_1979}%
  \BibitemOpen
  \bibfield  {author} {\bibinfo {author} {\bibfnamefont {P.~A.}\ \bibnamefont {Fedders}},\ }\href {https://doi.org/10.1103/PhysRevB.20.2588} {\bibfield  {journal} {\bibinfo  {journal} {Physical Review B}\ }\textbf {\bibinfo {volume} {20}},\ \bibinfo {pages} {2588} (\bibinfo {year} {1979})}\BibitemShut {NoStop}%
\bibitem [{\citenamefont {Ebner}\ and\ \citenamefont {Myles}(1975)}]{ebner_theoretical_1975}%
  \BibitemOpen
  \bibfield  {author} {\bibinfo {author} {\bibfnamefont {C.}~\bibnamefont {Ebner}}\ and\ \bibinfo {author} {\bibfnamefont {C.~W.}\ \bibnamefont {Myles}},\ }\href {https://doi.org/10.1103/PhysRevB.12.1638} {\bibfield  {journal} {\bibinfo  {journal} {Physical Review B}\ }\textbf {\bibinfo {volume} {12}},\ \bibinfo {pages} {1638} (\bibinfo {year} {1975})}\BibitemShut {NoStop}%
\bibitem [{\citenamefont {Fedders}(1986)}]{fedders_low-temperature_1986}%
  \BibitemOpen
  \bibfield  {author} {\bibinfo {author} {\bibfnamefont {P.~A.}\ \bibnamefont {Fedders}},\ }\href {https://doi.org/10.1103/PhysRevB.34.7489} {\bibfield  {journal} {\bibinfo  {journal} {Physical Review B}\ }\textbf {\bibinfo {volume} {34}},\ \bibinfo {pages} {7489} (\bibinfo {year} {1986})}\BibitemShut {NoStop}%
\bibitem [{\citenamefont {Fedders}(1975)}]{fedders_spin-resonance_1975}%
  \BibitemOpen
  \bibfield  {author} {\bibinfo {author} {\bibfnamefont {P.~A.}\ \bibnamefont {Fedders}},\ }\href {https://doi.org/10.1103/PhysRevB.11.995} {\bibfield  {journal} {\bibinfo  {journal} {Physical Review B}\ }\textbf {\bibinfo {volume} {11}},\ \bibinfo {pages} {995} (\bibinfo {year} {1975})}\BibitemShut {NoStop}%
\bibitem [{\citenamefont {Abragam}\ and\ \citenamefont {Carr}(1961)}]{abragam_principles_1961}%
  \BibitemOpen
  \bibfield  {author} {\bibinfo {author} {\bibfnamefont {A.}~\bibnamefont {Abragam}}\ and\ \bibinfo {author} {\bibfnamefont {H.~Y.}\ \bibnamefont {Carr}},\ }\href {https://doi.org/10.1063/1.3057238} {\bibfield  {journal} {\bibinfo  {journal} {Physics Today}\ }\textbf {\bibinfo {volume} {14}},\ \bibinfo {pages} {56} (\bibinfo {year} {1961})}\BibitemShut {NoStop}%
\bibitem [{\citenamefont {Scott}\ and\ \citenamefont {Jeffries}(1962)}]{scott_spin-lattice_1962}%
  \BibitemOpen
  \bibfield  {author} {\bibinfo {author} {\bibfnamefont {P.~L.}\ \bibnamefont {Scott}}\ and\ \bibinfo {author} {\bibfnamefont {C.~D.}\ \bibnamefont {Jeffries}},\ }\href {https://doi.org/10.1103/PhysRev.127.32} {\bibfield  {journal} {\bibinfo  {journal} {Physical Review}\ }\textbf {\bibinfo {volume} {127}},\ \bibinfo {pages} {32} (\bibinfo {year} {1962})}\BibitemShut {NoStop}%
\bibitem [{\citenamefont {Orbach}(1961)}]{orbach_spin-lattice_1961}%
  \BibitemOpen
  \bibfield  {author} {\bibinfo {author} {\bibfnamefont {R.~.}\ \bibnamefont {Orbach}},\ }\href {https://doi.org/10.1098/rspa.1961.0211} {\bibfield  {journal} {\bibinfo  {journal} {Proceedings of the Royal Society of London. Series A. Mathematical and Physical Sciences}\ }\textbf {\bibinfo {volume} {264}},\ \bibinfo {pages} {458} (\bibinfo {year} {1961})}\BibitemShut {NoStop}%
\bibitem [{\citenamefont {Hjorth~Larsen}\ \emph {et~al.}(2017)\citenamefont {Hjorth~Larsen}, \citenamefont {J{\o}rgen~Mortensen}, \citenamefont {Blomqvist}, \citenamefont {Castelli}, \citenamefont {Christensen}, \citenamefont {Du{\l}ak}, \citenamefont {Friis}, \citenamefont {Groves}, \citenamefont {Hammer}, \citenamefont {Hargus}, \citenamefont {Hermes}, \citenamefont {Jennings}, \citenamefont {Bjerre~Jensen}, \citenamefont {Kermode}, \citenamefont {Kitchin}, \citenamefont {Leonhard~Kolsbjerg}, \citenamefont {Kubal}, \citenamefont {Kaasbjerg}, \citenamefont {Lysgaard}, \citenamefont {Bergmann~Maronsson}, \citenamefont {Maxson}, \citenamefont {Olsen}, \citenamefont {Pastewka}, \citenamefont {Peterson}, \citenamefont {Rostgaard}, \citenamefont {Schi{\o}tz}, \citenamefont {Sch{\"u}tt}, \citenamefont {Strange}, \citenamefont {Thygesen}, \citenamefont {Vegge}, \citenamefont {Vilhelmsen}, \citenamefont {Walter}, \citenamefont {Zeng},\ and\ \citenamefont {Jacobsen}}]{ASE}%
  \BibitemOpen
  \bibfield  {author} {\bibinfo {author} {\bibfnamefont {A.}~\bibnamefont {Hjorth~Larsen}}, \bibinfo {author} {\bibfnamefont {J.}~\bibnamefont {J{\o}rgen~Mortensen}}, \bibinfo {author} {\bibfnamefont {J.}~\bibnamefont {Blomqvist}}, \bibinfo {author} {\bibfnamefont {I.~E.}\ \bibnamefont {Castelli}}, \bibinfo {author} {\bibfnamefont {R.}~\bibnamefont {Christensen}}, \bibinfo {author} {\bibfnamefont {M.}~\bibnamefont {Du{\l}ak}}, \bibinfo {author} {\bibfnamefont {J.}~\bibnamefont {Friis}}, \bibinfo {author} {\bibfnamefont {M.~N.}\ \bibnamefont {Groves}}, \bibinfo {author} {\bibfnamefont {B.}~\bibnamefont {Hammer}}, \bibinfo {author} {\bibfnamefont {C.}~\bibnamefont {Hargus}}, \bibinfo {author} {\bibfnamefont {E.~D.}\ \bibnamefont {Hermes}}, \bibinfo {author} {\bibfnamefont {P.~C.}\ \bibnamefont {Jennings}}, \bibinfo {author} {\bibfnamefont {P.}~\bibnamefont {Bjerre~Jensen}}, \bibinfo {author} {\bibfnamefont {J.}~\bibnamefont {Kermode}}, \bibinfo {author} {\bibfnamefont {J.~R.}\ \bibnamefont {Kitchin}}, \bibinfo
  {author} {\bibfnamefont {E.}~\bibnamefont {Leonhard~Kolsbjerg}}, \bibinfo {author} {\bibfnamefont {J.}~\bibnamefont {Kubal}}, \bibinfo {author} {\bibfnamefont {K.}~\bibnamefont {Kaasbjerg}}, \bibinfo {author} {\bibfnamefont {S.}~\bibnamefont {Lysgaard}}, \bibinfo {author} {\bibfnamefont {J.}~\bibnamefont {Bergmann~Maronsson}}, \bibinfo {author} {\bibfnamefont {T.}~\bibnamefont {Maxson}}, \bibinfo {author} {\bibfnamefont {T.}~\bibnamefont {Olsen}}, \bibinfo {author} {\bibfnamefont {L.}~\bibnamefont {Pastewka}}, \bibinfo {author} {\bibfnamefont {A.}~\bibnamefont {Peterson}}, \bibinfo {author} {\bibfnamefont {C.}~\bibnamefont {Rostgaard}}, \bibinfo {author} {\bibfnamefont {J.}~\bibnamefont {Schi{\o}tz}}, \bibinfo {author} {\bibfnamefont {O.}~\bibnamefont {Sch{\"u}tt}}, \bibinfo {author} {\bibfnamefont {M.}~\bibnamefont {Strange}}, \bibinfo {author} {\bibfnamefont {K.~S.}\ \bibnamefont {Thygesen}}, \bibinfo {author} {\bibfnamefont {T.}~\bibnamefont {Vegge}}, \bibinfo {author} {\bibfnamefont {L.}~\bibnamefont
  {Vilhelmsen}}, \bibinfo {author} {\bibfnamefont {M.}~\bibnamefont {Walter}}, \bibinfo {author} {\bibfnamefont {Z.}~\bibnamefont {Zeng}},\ and\ \bibinfo {author} {\bibfnamefont {K.~W.}\ \bibnamefont {Jacobsen}},\ }\href {https://doi.org/10.1088/1361-648X/aa680e} {\bibfield  {journal} {\bibinfo  {journal} {J. Phys.: Condens. Matter}\ }\textbf {\bibinfo {volume} {29}},\ \bibinfo {pages} {273002} (\bibinfo {year} {2017})}\BibitemShut {NoStop}%
\bibitem [{\citenamefont {Neumann}\ \emph {et~al.}(2024)\citenamefont {Neumann}, \citenamefont {Gin}, \citenamefont {Rhodes}, \citenamefont {Bennett}, \citenamefont {Li}, \citenamefont {Choubisa}, \citenamefont {Hussey},\ and\ \citenamefont {Godwin}}]{Orb}%
  \BibitemOpen
  \bibfield  {author} {\bibinfo {author} {\bibfnamefont {M.}~\bibnamefont {Neumann}}, \bibinfo {author} {\bibfnamefont {J.}~\bibnamefont {Gin}}, \bibinfo {author} {\bibfnamefont {B.}~\bibnamefont {Rhodes}}, \bibinfo {author} {\bibfnamefont {S.}~\bibnamefont {Bennett}}, \bibinfo {author} {\bibfnamefont {Z.}~\bibnamefont {Li}}, \bibinfo {author} {\bibfnamefont {H.}~\bibnamefont {Choubisa}}, \bibinfo {author} {\bibfnamefont {A.}~\bibnamefont {Hussey}},\ and\ \bibinfo {author} {\bibfnamefont {J.}~\bibnamefont {Godwin}},\ }\href {https://doi.org/10.48550/arXiv.2410.22570} {\bibinfo {title} {Orb: {A} fast, scalable neural network potential}} (\bibinfo {year} {2024}),\ \Eprint {https://arxiv.org/abs/2410.22570} {arXiv:2410.22570 [cond-mat.mtrl-sci]} \BibitemShut {NoStop}%
\bibitem [{\citenamefont {Eastman}\ \emph {et~al.}(2026)\citenamefont {Eastman}, \citenamefont {Pretti},\ and\ \citenamefont {Markland}}]{MarklandOrb}%
  \BibitemOpen
  \bibfield  {author} {\bibinfo {author} {\bibfnamefont {P.}~\bibnamefont {Eastman}}, \bibinfo {author} {\bibfnamefont {E.}~\bibnamefont {Pretti}},\ and\ \bibinfo {author} {\bibfnamefont {T.~E.}\ \bibnamefont {Markland}},\ }\bibfield  {journal} {\bibinfo  {journal} {J. Chem. Theory Comput.}\ }\href {https://doi.org/10.1021/acs.jctc.6c00130} {10.1021/acs.jctc.6c00130} (\bibinfo {year} {2026})\BibitemShut {NoStop}%
\bibitem [{\citenamefont {Swendsen}\ and\ \citenamefont {Wang}(1986)}]{swendsen_replica_1986}%
  \BibitemOpen
  \bibfield  {author} {\bibinfo {author} {\bibfnamefont {R.~H.}\ \bibnamefont {Swendsen}}\ and\ \bibinfo {author} {\bibfnamefont {J.-S.}\ \bibnamefont {Wang}},\ }\href {https://doi.org/10.1103/PhysRevLett.57.2607} {\bibfield  {journal} {\bibinfo  {journal} {Phys. Rev. Lett.}\ }\textbf {\bibinfo {volume} {57}},\ \bibinfo {pages} {2607} (\bibinfo {year} {1986})}\BibitemShut {NoStop}%
\bibitem [{\citenamefont {Hukushima}\ and\ \citenamefont {Nemoto}(1996)}]{hukushima_exchange_1996}%
  \BibitemOpen
  \bibfield  {author} {\bibinfo {author} {\bibfnamefont {K.}~\bibnamefont {Hukushima}}\ and\ \bibinfo {author} {\bibfnamefont {K.}~\bibnamefont {Nemoto}},\ }\href {https://doi.org/10.1143/JPSJ.65.1604} {\bibfield  {journal} {\bibinfo  {journal} {J. Phys. Soc. Jpn.}\ }\textbf {\bibinfo {volume} {65}},\ \bibinfo {pages} {1604} (\bibinfo {year} {1996})}\BibitemShut {NoStop}%
\bibitem [{\citenamefont {Sugita}\ and\ \citenamefont {Okamoto}(1999)}]{sugita_replica_1999}%
  \BibitemOpen
  \bibfield  {author} {\bibinfo {author} {\bibfnamefont {Y.}~\bibnamefont {Sugita}}\ and\ \bibinfo {author} {\bibfnamefont {Y.}~\bibnamefont {Okamoto}},\ }\href {https://doi.org/10.1016/S0009-2614(99)01123-9} {\bibfield  {journal} {\bibinfo  {journal} {Chem. Phys. Lett.}\ }\textbf {\bibinfo {volume} {314}},\ \bibinfo {pages} {141} (\bibinfo {year} {1999})}\BibitemShut {NoStop}%
\bibitem [{\citenamefont {Grigera}\ and\ \citenamefont {Parisi}(2001)}]{grigera_fast_2001}%
  \BibitemOpen
  \bibfield  {author} {\bibinfo {author} {\bibfnamefont {T.~S.}\ \bibnamefont {Grigera}}\ and\ \bibinfo {author} {\bibfnamefont {G.}~\bibnamefont {Parisi}},\ }\href {https://doi.org/10.1103/PhysRevE.63.045102} {\bibfield  {journal} {\bibinfo  {journal} {Phys. Rev. E}\ }\textbf {\bibinfo {volume} {63}},\ \bibinfo {pages} {045102} (\bibinfo {year} {2001})}\BibitemShut {NoStop}%
\bibitem [{\citenamefont {Ninarello}\ \emph {et~al.}(2017)\citenamefont {Ninarello}, \citenamefont {Berthier},\ and\ \citenamefont {Coslovich}}]{ninarello_models_2017}%
  \BibitemOpen
  \bibfield  {author} {\bibinfo {author} {\bibfnamefont {A.}~\bibnamefont {Ninarello}}, \bibinfo {author} {\bibfnamefont {L.}~\bibnamefont {Berthier}},\ and\ \bibinfo {author} {\bibfnamefont {D.}~\bibnamefont {Coslovich}},\ }\href {https://doi.org/10.1103/PhysRevX.7.021039} {\bibfield  {journal} {\bibinfo  {journal} {Phys. Rev. X}\ }\textbf {\bibinfo {volume} {7}},\ \bibinfo {pages} {021039} (\bibinfo {year} {2017})}\BibitemShut {NoStop}%
\bibitem [{SM()}]{SM}%
  \BibitemOpen
  \href@noop {} {}\bibinfo {note} {See Supplemental Material at [URL will be inserted by publisher] for details on the experimental spectra and kinetics for all H$_2$:CO$_2$ and H$_2$:N$_2$O conditions and Orb neural network validation includes Refs. [63--72].}\BibitemShut {Stop}%
\bibitem [{\citenamefont {Escribano}\ \emph {et~al.}(2013)\citenamefont {Escribano}, \citenamefont {Muñoz~Caro}, \citenamefont {Cruz-Diaz}, \citenamefont {Rodríguez-Lazcano},\ and\ \citenamefont {Maté}}]{escribano_crystallization_2013}%
  \BibitemOpen
  \bibfield  {author} {\bibinfo {author} {\bibfnamefont {R.~M.}\ \bibnamefont {Escribano}}, \bibinfo {author} {\bibfnamefont {G.~M.}\ \bibnamefont {Muñoz~Caro}}, \bibinfo {author} {\bibfnamefont {G.~A.}\ \bibnamefont {Cruz-Diaz}}, \bibinfo {author} {\bibfnamefont {Y.}~\bibnamefont {Rodríguez-Lazcano}},\ and\ \bibinfo {author} {\bibfnamefont {B.}~\bibnamefont {Maté}},\ }\href {https://doi.org/10.1073/pnas.1222228110} {\bibfield  {journal} {\bibinfo  {journal} {Proceedings of the National Academy of Sciences}\ }\textbf {\bibinfo {volume} {110}},\ \bibinfo {pages} {12899} (\bibinfo {year} {2013})}\BibitemShut {NoStop}%
\bibitem [{\citenamefont {Bini}\ \emph {et~al.}(1992)\citenamefont {Bini}, \citenamefont {Jodl},\ and\ \citenamefont {Califano}}]{bini_relaxation_1992}%
  \BibitemOpen
  \bibfield  {author} {\bibinfo {author} {\bibfnamefont {R.}~\bibnamefont {Bini}}, \bibinfo {author} {\bibfnamefont {H.~J.}\ \bibnamefont {Jodl}},\ and\ \bibinfo {author} {\bibfnamefont {S.}~\bibnamefont {Califano}},\ }\href {https://doi.org/10.1103/PhysRevB.45.5244} {\bibfield  {journal} {\bibinfo  {journal} {Physical Review B}\ }\textbf {\bibinfo {volume} {45}},\ \bibinfo {pages} {5244} (\bibinfo {year} {1992})}\BibitemShut {NoStop}%
\bibitem [{\citenamefont {Buckingham}(1959)}]{buckingham_molecular_1959}%
  \BibitemOpen
  \bibfield  {author} {\bibinfo {author} {\bibfnamefont {A.~D.}\ \bibnamefont {Buckingham}},\ }\href {https://doi.org/10.1039/qr9591300183} {\bibfield  {journal} {\bibinfo  {journal} {Quarterly Reviews, Chemical Society}\ }\textbf {\bibinfo {volume} {13}},\ \bibinfo {pages} {183} (\bibinfo {year} {1959})}\BibitemShut {NoStop}%
\bibitem [{\citenamefont {Wilson}\ \emph {et~al.}(1981)\citenamefont {Wilson}, \citenamefont {Bisson},\ and\ \citenamefont {Baskes}}]{wilson_self-trapping_1981}%
  \BibitemOpen
  \bibfield  {author} {\bibinfo {author} {\bibfnamefont {W.~D.}\ \bibnamefont {Wilson}}, \bibinfo {author} {\bibfnamefont {C.~L.}\ \bibnamefont {Bisson}},\ and\ \bibinfo {author} {\bibfnamefont {M.~I.}\ \bibnamefont {Baskes}},\ }\href {https://doi.org/10.1103/PhysRevB.24.5616} {\bibfield  {journal} {\bibinfo  {journal} {Phys. Rev. B}\ }\textbf {\bibinfo {volume} {24}},\ \bibinfo {pages} {5616} (\bibinfo {year} {1981})}\BibitemShut {NoStop}%
\bibitem [{\citenamefont {Trinkaus}(1983)}]{trinkaus_energetics_1983}%
  \BibitemOpen
  \bibfield  {author} {\bibinfo {author} {\bibfnamefont {H.}~\bibnamefont {Trinkaus}},\ }\href {https://doi.org/10.1080/00337578308207371} {\bibfield  {journal} {\bibinfo  {journal} {Radiat. Eff.}\ }\textbf {\bibinfo {volume} {78}},\ \bibinfo {pages} {189} (\bibinfo {year} {1983})}\BibitemShut {NoStop}%
\bibitem [{\citenamefont {Trinkaus}\ and\ \citenamefont {Singh}(2003)}]{trinkaus_helium_2003}%
  \BibitemOpen
  \bibfield  {author} {\bibinfo {author} {\bibfnamefont {H.}~\bibnamefont {Trinkaus}}\ and\ \bibinfo {author} {\bibfnamefont {B.~N.}\ \bibnamefont {Singh}},\ }\href {https://doi.org/10.1016/j.jnucmat.2003.09.001} {\bibfield  {journal} {\bibinfo  {journal} {J. Nucl. Mater.}\ }\textbf {\bibinfo {volume} {323}},\ \bibinfo {pages} {229} (\bibinfo {year} {2003})}\BibitemShut {NoStop}%
\bibitem [{\citenamefont {Zhou}\ \emph {et~al.}(2021)\citenamefont {Zhou}, \citenamefont {Bartelt},\ and\ \citenamefont {Sills}}]{zhou_enabling_2021}%
  \BibitemOpen
  \bibfield  {author} {\bibinfo {author} {\bibfnamefont {X.~W.}\ \bibnamefont {Zhou}}, \bibinfo {author} {\bibfnamefont {N.~C.}\ \bibnamefont {Bartelt}},\ and\ \bibinfo {author} {\bibfnamefont {R.~B.}\ \bibnamefont {Sills}},\ }\href {https://doi.org/10.1103/PhysRevB.103.014108} {\bibfield  {journal} {\bibinfo  {journal} {Phys. Rev. B}\ }\textbf {\bibinfo {volume} {103}},\ \bibinfo {pages} {014108} (\bibinfo {year} {2021})}\BibitemShut {NoStop}%
\bibitem [{\citenamefont {Falk}\ and\ \citenamefont {Langer}(1998)}]{falk_dynamics_1998}%
  \BibitemOpen
  \bibfield  {author} {\bibinfo {author} {\bibfnamefont {M.~L.}\ \bibnamefont {Falk}}\ and\ \bibinfo {author} {\bibfnamefont {J.~S.}\ \bibnamefont {Langer}},\ }\href {https://doi.org/10.1103/PhysRevE.57.7192} {\bibfield  {journal} {\bibinfo  {journal} {Phys. Rev. E}\ }\textbf {\bibinfo {volume} {57}},\ \bibinfo {pages} {7192} (\bibinfo {year} {1998})}\BibitemShut {NoStop}%
\bibitem [{\citenamefont {Shimizu}\ \emph {et~al.}(2007)\citenamefont {Shimizu}, \citenamefont {Ogata},\ and\ \citenamefont {Li}}]{shimizu_theory_2007}%
  \BibitemOpen
  \bibfield  {author} {\bibinfo {author} {\bibfnamefont {F.}~\bibnamefont {Shimizu}}, \bibinfo {author} {\bibfnamefont {S.}~\bibnamefont {Ogata}},\ and\ \bibinfo {author} {\bibfnamefont {J.}~\bibnamefont {Li}},\ }\href {https://doi.org/10.2320/matertrans.MJ200769} {\bibfield  {journal} {\bibinfo  {journal} {Mater. Trans.}\ }\textbf {\bibinfo {volume} {48}},\ \bibinfo {pages} {2923} (\bibinfo {year} {2007})}\BibitemShut {NoStop}%
\bibitem [{\citenamefont {Warren}\ \emph {et~al.}(1980)\citenamefont {Warren}, \citenamefont {Smith},\ and\ \citenamefont {Guillory}}]{warren_infrared_1980}%
  \BibitemOpen
  \bibfield  {author} {\bibinfo {author} {\bibfnamefont {J.~A.}\ \bibnamefont {Warren}}, \bibinfo {author} {\bibfnamefont {G.~R.}\ \bibnamefont {Smith}},\ and\ \bibinfo {author} {\bibfnamefont {W.~A.}\ \bibnamefont {Guillory}},\ }\href {https://doi.org/10.1063/1.439774} {\bibfield  {journal} {\bibinfo  {journal} {The Journal of Chemical Physics}\ }\textbf {\bibinfo {volume} {72}},\ \bibinfo {pages} {4901} (\bibinfo {year} {1980})}\BibitemShut {NoStop}%
\bibitem [{\citenamefont {Mamone}\ \emph {et~al.}(2009)\citenamefont {Mamone}, \citenamefont {Ge}, \citenamefont {Hüvonen}, \citenamefont {Nagel}, \citenamefont {Danquigny}, \citenamefont {Cuda}, \citenamefont {Grossel}, \citenamefont {Murata}, \citenamefont {Komatsu}, \citenamefont {Levitt}, \citenamefont {Rõõm},\ and\ \citenamefont {Carravetta}}]{mamone_rotor_2009}%
  \BibitemOpen
  \bibfield  {author} {\bibinfo {author} {\bibfnamefont {S.}~\bibnamefont {Mamone}}, \bibinfo {author} {\bibfnamefont {M.}~\bibnamefont {Ge}}, \bibinfo {author} {\bibfnamefont {D.}~\bibnamefont {Hüvonen}}, \bibinfo {author} {\bibfnamefont {U.}~\bibnamefont {Nagel}}, \bibinfo {author} {\bibfnamefont {A.}~\bibnamefont {Danquigny}}, \bibinfo {author} {\bibfnamefont {F.}~\bibnamefont {Cuda}}, \bibinfo {author} {\bibfnamefont {M.~C.}\ \bibnamefont {Grossel}}, \bibinfo {author} {\bibfnamefont {Y.}~\bibnamefont {Murata}}, \bibinfo {author} {\bibfnamefont {K.}~\bibnamefont {Komatsu}}, \bibinfo {author} {\bibfnamefont {M.~H.}\ \bibnamefont {Levitt}}, \bibinfo {author} {\bibfnamefont {T.}~\bibnamefont {Rõõm}},\ and\ \bibinfo {author} {\bibfnamefont {M.}~\bibnamefont {Carravetta}},\ }\href {https://doi.org/10.1063/1.3080163} {\bibfield  {journal} {\bibinfo  {journal} {The Journal of Chemical Physics}\ }\textbf {\bibinfo {volume} {130}},\ \bibinfo {pages} {081103} (\bibinfo {year} {2009})}\BibitemShut {NoStop}%
\bibitem [{\citenamefont {Mamone}\ \emph {et~al.}(2011)\citenamefont {Mamone}, \citenamefont {Chen}, \citenamefont {Bhattacharyya}, \citenamefont {Levitt}, \citenamefont {Lawler}, \citenamefont {Horsewill}, \citenamefont {Rõõm}, \citenamefont {Bačić},\ and\ \citenamefont {Turro}}]{mamone_theory_2011}%
  \BibitemOpen
  \bibfield  {author} {\bibinfo {author} {\bibfnamefont {S.}~\bibnamefont {Mamone}}, \bibinfo {author} {\bibfnamefont {J.~Y.-C.}\ \bibnamefont {Chen}}, \bibinfo {author} {\bibfnamefont {R.}~\bibnamefont {Bhattacharyya}}, \bibinfo {author} {\bibfnamefont {M.~H.}\ \bibnamefont {Levitt}}, \bibinfo {author} {\bibfnamefont {R.~G.}\ \bibnamefont {Lawler}}, \bibinfo {author} {\bibfnamefont {A.~J.}\ \bibnamefont {Horsewill}}, \bibinfo {author} {\bibfnamefont {T.}~\bibnamefont {Rõõm}}, \bibinfo {author} {\bibfnamefont {Z.}~\bibnamefont {Bačić}},\ and\ \bibinfo {author} {\bibfnamefont {N.~J.}\ \bibnamefont {Turro}},\ }\href {https://doi.org/10.1016/j.ccr.2010.12.029} {\bibfield  {journal} {\bibinfo  {journal} {Coordination Chemistry Reviews}\ }\textbf {\bibinfo {volume} {255}},\ \bibinfo {pages} {938} (\bibinfo {year} {2011})}\BibitemShut {NoStop}%
\bibitem [{\citenamefont {Hixson}\ \emph {et~al.}(1992)\citenamefont {Hixson}, \citenamefont {Wojcik}, \citenamefont {Devlin}, \citenamefont {Devlin},\ and\ \citenamefont {Buch}}]{hixson_experimental_1992}%
  \BibitemOpen
  \bibfield  {author} {\bibinfo {author} {\bibfnamefont {H.~G.}\ \bibnamefont {Hixson}}, \bibinfo {author} {\bibfnamefont {M.~J.}\ \bibnamefont {Wojcik}}, \bibinfo {author} {\bibfnamefont {M.~S.}\ \bibnamefont {Devlin}}, \bibinfo {author} {\bibfnamefont {J.~P.}\ \bibnamefont {Devlin}},\ and\ \bibinfo {author} {\bibfnamefont {V.}~\bibnamefont {Buch}},\ }\href {https://doi.org/10.1063/1.463240} {\bibfield  {journal} {\bibinfo  {journal} {The Journal of Chemical Physics}\ }\textbf {\bibinfo {volume} {97}},\ \bibinfo {pages} {753} (\bibinfo {year} {1992})}\BibitemShut {NoStop}%
\bibitem [{\citenamefont {Sugimoto}\ and\ \citenamefont {Fukutani}(2011)}]{sugimoto_electric-field-induced_2011}%
  \BibitemOpen
  \bibfield  {author} {\bibinfo {author} {\bibfnamefont {T.}~\bibnamefont {Sugimoto}}\ and\ \bibinfo {author} {\bibfnamefont {K.}~\bibnamefont {Fukutani}},\ }\href {https://doi.org/10.1038/nphys1883} {\bibfield  {journal} {\bibinfo  {journal} {Nature Physics}\ }\textbf {\bibinfo {volume} {7}},\ \bibinfo {pages} {307} (\bibinfo {year} {2011})}\BibitemShut {NoStop}%
\bibitem [{\citenamefont {Yamakawa}\ \emph {et~al.}(2020)\citenamefont {Yamakawa}, \citenamefont {Ishibashi}, \citenamefont {Namiyoshi}, \citenamefont {Azuma},\ and\ \citenamefont {Arakawa}}]{yamakawa_fast_2020}%
  \BibitemOpen
  \bibfield  {author} {\bibinfo {author} {\bibfnamefont {K.}~\bibnamefont {Yamakawa}}, \bibinfo {author} {\bibfnamefont {A.}~\bibnamefont {Ishibashi}}, \bibinfo {author} {\bibfnamefont {T.}~\bibnamefont {Namiyoshi}}, \bibinfo {author} {\bibfnamefont {Y.}~\bibnamefont {Azuma}},\ and\ \bibinfo {author} {\bibfnamefont {I.}~\bibnamefont {Arakawa}},\ }\href {https://doi.org/10.1103/PhysRevB.102.041401} {\bibfield  {journal} {\bibinfo  {journal} {Physical Review B}\ }\textbf {\bibinfo {volume} {102}},\ \bibinfo {pages} {041401} (\bibinfo {year} {2020})}\BibitemShut {NoStop}%
\bibitem [{dat()}]{data_availability}%
  \BibitemOpen
  \href@noop {} {}\bibinfo {note} {The data that support the findings of this study are available in Figshare at 10.6084/m9.figshare.34930377.}\BibitemShut {Stop}%
\bibitem [{\citenamefont {Rhodes}\ \emph {et~al.}(2025)\citenamefont {Rhodes}, \citenamefont {Vandenhaute}, \citenamefont {{\v{S}}imkus}, \citenamefont {Gin}, \citenamefont {Godwin}, \citenamefont {Duignan},\ and\ \citenamefont {Neumann}}]{orbv3}%
  \BibitemOpen
  \bibfield  {author} {\bibinfo {author} {\bibfnamefont {B.}~\bibnamefont {Rhodes}}, \bibinfo {author} {\bibfnamefont {S.}~\bibnamefont {Vandenhaute}}, \bibinfo {author} {\bibfnamefont {V.}~\bibnamefont {{\v{S}}imkus}}, \bibinfo {author} {\bibfnamefont {J.}~\bibnamefont {Gin}}, \bibinfo {author} {\bibfnamefont {J.}~\bibnamefont {Godwin}}, \bibinfo {author} {\bibfnamefont {T.}~\bibnamefont {Duignan}},\ and\ \bibinfo {author} {\bibfnamefont {M.}~\bibnamefont {Neumann}},\ }\bibfield  {journal} {\bibinfo  {journal} {arXiv preprint arXiv:2504.06231}\ }\href {https://doi.org/10.48550/arXiv.2504.06231} {10.48550/arXiv.2504.06231} (\bibinfo {year} {2025}),\ \Eprint {https://arxiv.org/abs/2504.06231} {arXiv:2504.06231} \BibitemShut {NoStop}%
\bibitem [{\citenamefont {Levine}\ \emph {et~al.}(2025)\citenamefont {Levine}, \citenamefont {Shuaibi}, \citenamefont {Spotte-Smith}, \citenamefont {Taylor}, \citenamefont {Hasyim}, \citenamefont {Michel}, \citenamefont {Batatia}, \citenamefont {Cs{\'a}nyi}, \citenamefont {Dzamba}, \citenamefont {Eastman}, \citenamefont {Frey}, \citenamefont {Fu}, \citenamefont {Gharakhanyan}, \citenamefont {Krishnapriyan}, \citenamefont {Rackers}, \citenamefont {Raja}, \citenamefont {Rizvi}, \citenamefont {Rosen}, \citenamefont {Ulissi}, \citenamefont {Vargas}, \citenamefont {Zitnick}, \citenamefont {Blau},\ and\ \citenamefont {Wood}}]{omol25}%
  \BibitemOpen
  \bibfield  {author} {\bibinfo {author} {\bibfnamefont {D.~S.}\ \bibnamefont {Levine}}, \bibinfo {author} {\bibfnamefont {M.}~\bibnamefont {Shuaibi}}, \bibinfo {author} {\bibfnamefont {E.~W.~C.}\ \bibnamefont {Spotte-Smith}}, \bibinfo {author} {\bibfnamefont {M.~G.}\ \bibnamefont {Taylor}}, \bibinfo {author} {\bibfnamefont {M.~R.}\ \bibnamefont {Hasyim}}, \bibinfo {author} {\bibfnamefont {K.}~\bibnamefont {Michel}}, \bibinfo {author} {\bibfnamefont {I.}~\bibnamefont {Batatia}}, \bibinfo {author} {\bibfnamefont {G.}~\bibnamefont {Cs{\'a}nyi}}, \bibinfo {author} {\bibfnamefont {M.}~\bibnamefont {Dzamba}}, \bibinfo {author} {\bibfnamefont {P.}~\bibnamefont {Eastman}}, \bibinfo {author} {\bibfnamefont {N.~C.}\ \bibnamefont {Frey}}, \bibinfo {author} {\bibfnamefont {X.}~\bibnamefont {Fu}}, \bibinfo {author} {\bibfnamefont {V.}~\bibnamefont {Gharakhanyan}}, \bibinfo {author} {\bibfnamefont {A.~S.}\ \bibnamefont {Krishnapriyan}}, \bibinfo {author} {\bibfnamefont {J.~A.}\ \bibnamefont {Rackers}}, \bibinfo {author}
  {\bibfnamefont {S.}~\bibnamefont {Raja}}, \bibinfo {author} {\bibfnamefont {A.}~\bibnamefont {Rizvi}}, \bibinfo {author} {\bibfnamefont {A.~S.}\ \bibnamefont {Rosen}}, \bibinfo {author} {\bibfnamefont {Z.}~\bibnamefont {Ulissi}}, \bibinfo {author} {\bibfnamefont {S.}~\bibnamefont {Vargas}}, \bibinfo {author} {\bibfnamefont {C.~L.}\ \bibnamefont {Zitnick}}, \bibinfo {author} {\bibfnamefont {S.~M.}\ \bibnamefont {Blau}},\ and\ \bibinfo {author} {\bibfnamefont {B.~M.}\ \bibnamefont {Wood}},\ }\bibfield  {journal} {\bibinfo  {journal} {arXiv preprint arXiv:2505.08762}\ }\href {https://doi.org/10.48550/arXiv.2505.08762} {10.48550/arXiv.2505.08762} (\bibinfo {year} {2025}),\ \Eprint {https://arxiv.org/abs/2505.08762} {arXiv:2505.08762} \BibitemShut {NoStop}%
\bibitem [{\citenamefont {Mardirossian}\ and\ \citenamefont {Head-Gordon}(2016)}]{wb97mv}%
  \BibitemOpen
  \bibfield  {author} {\bibinfo {author} {\bibfnamefont {N.}~\bibnamefont {Mardirossian}}\ and\ \bibinfo {author} {\bibfnamefont {M.}~\bibnamefont {Head-Gordon}},\ }\href {https://doi.org/10.1063/1.4952647} {\bibfield  {journal} {\bibinfo  {journal} {J. Chem. Phys.}\ }\textbf {\bibinfo {volume} {144}},\ \bibinfo {pages} {214110} (\bibinfo {year} {2016})}\BibitemShut {NoStop}%
\bibitem [{\citenamefont {Vydrov}\ and\ \citenamefont {Van~Voorhis}(2010)}]{vv10}%
  \BibitemOpen
  \bibfield  {author} {\bibinfo {author} {\bibfnamefont {O.~A.}\ \bibnamefont {Vydrov}}\ and\ \bibinfo {author} {\bibfnamefont {T.}~\bibnamefont {Van~Voorhis}},\ }\href {https://doi.org/10.1063/1.3521275} {\bibfield  {journal} {\bibinfo  {journal} {J. Chem. Phys.}\ }\textbf {\bibinfo {volume} {133}},\ \bibinfo {pages} {244103} (\bibinfo {year} {2010})}\BibitemShut {NoStop}%
\bibitem [{\citenamefont {Weigend}\ and\ \citenamefont {Ahlrichs}(2005)}]{def2}%
  \BibitemOpen
  \bibfield  {author} {\bibinfo {author} {\bibfnamefont {F.}~\bibnamefont {Weigend}}\ and\ \bibinfo {author} {\bibfnamefont {R.}~\bibnamefont {Ahlrichs}},\ }\href {https://doi.org/10.1039/b508541a} {\bibfield  {journal} {\bibinfo  {journal} {Phys. Chem. Chem. Phys.}\ }\textbf {\bibinfo {volume} {7}},\ \bibinfo {pages} {3297} (\bibinfo {year} {2005})}\BibitemShut {NoStop}%
\bibitem [{\citenamefont {Rappoport}\ and\ \citenamefont {Furche}(2010)}]{def2d}%
  \BibitemOpen
  \bibfield  {author} {\bibinfo {author} {\bibfnamefont {D.}~\bibnamefont {Rappoport}}\ and\ \bibinfo {author} {\bibfnamefont {F.}~\bibnamefont {Furche}},\ }\href {https://doi.org/10.1063/1.3484283} {\bibfield  {journal} {\bibinfo  {journal} {J. Chem. Phys.}\ }\textbf {\bibinfo {volume} {133}},\ \bibinfo {pages} {134105} (\bibinfo {year} {2010})}\BibitemShut {NoStop}%
\bibitem [{\citenamefont {Neese}\ \emph {et~al.}(2009)\citenamefont {Neese}, \citenamefont {Wennmohs}, \citenamefont {Hansen},\ and\ \citenamefont {Becker}}]{rijcosx}%
  \BibitemOpen
  \bibfield  {author} {\bibinfo {author} {\bibfnamefont {F.}~\bibnamefont {Neese}}, \bibinfo {author} {\bibfnamefont {F.}~\bibnamefont {Wennmohs}}, \bibinfo {author} {\bibfnamefont {A.}~\bibnamefont {Hansen}},\ and\ \bibinfo {author} {\bibfnamefont {U.}~\bibnamefont {Becker}},\ }\href {https://doi.org/10.1016/j.chemphys.2008.10.036} {\bibfield  {journal} {\bibinfo  {journal} {Chem. Phys.}\ }\textbf {\bibinfo {volume} {356}},\ \bibinfo {pages} {98} (\bibinfo {year} {2009})}\BibitemShut {NoStop}%
\bibitem [{\citenamefont {Weigend}(2006)}]{def2j}%
  \BibitemOpen
  \bibfield  {author} {\bibinfo {author} {\bibfnamefont {F.}~\bibnamefont {Weigend}},\ }\href {https://doi.org/10.1039/b515623h} {\bibfield  {journal} {\bibinfo  {journal} {Phys. Chem. Chem. Phys.}\ }\textbf {\bibinfo {volume} {8}},\ \bibinfo {pages} {1057} (\bibinfo {year} {2006})}\BibitemShut {NoStop}%
\bibitem [{\citenamefont {Neese}(2012)}]{orca2012}%
  \BibitemOpen
  \bibfield  {author} {\bibinfo {author} {\bibfnamefont {F.}~\bibnamefont {Neese}},\ }\href {https://doi.org/10.1002/wcms.81} {\bibfield  {journal} {\bibinfo  {journal} {WIREs Comput. Mol. Sci.}\ }\textbf {\bibinfo {volume} {2}},\ \bibinfo {pages} {73} (\bibinfo {year} {2012})}\BibitemShut {NoStop}%
\bibitem [{\citenamefont {Neese}(2025)}]{orca6}%
  \BibitemOpen
  \bibfield  {author} {\bibinfo {author} {\bibfnamefont {F.}~\bibnamefont {Neese}},\ }\href {https://doi.org/10.1002/wcms.70019} {\bibfield  {journal} {\bibinfo  {journal} {WIREs Comput. Mol. Sci.}\ }\textbf {\bibinfo {volume} {15}},\ \bibinfo {pages} {e70019} (\bibinfo {year} {2025})}\BibitemShut {NoStop}%
\end{thebibliography}
 % for arxiv submission

\end{document}